\documentclass[
twocolumn,showpacs,preprintnumbers,amsmath,amssymb]{revtex4}
\usepackage{graphicx}
\usepackage[english]{babel}
\usepackage{microtype}		
\usepackage[breaklinks=true,colorlinks=true,linkcolor=blue,urlcolor=blue,citecolor=blue]{hyperref}

\usepackage{xcolor}	
\usepackage{mathtools}
\usepackage{braket}
\usepackage{tikz}
 \usepackage{printlen}

 \usepackage{longtable}
  \usepackage{booktabs}
 \usepackage{siunitx}
 \usepackage{csquotes}

\newcommand{\op}[1]{\ensuremath{\hat{#1}}}
\newcommand{\ladderdown}{\ensuremath{\op{a}^{\vphantom{\dagger}}}}
\newcommand{\ladderup}{\ensuremath{\op{a}^\dagger}}
\newcommand{\annihilop}{\ladderdown}
\newcommand{\creationop}{\ladderup}

\begin{document}
\bibliographystyle{apsrev}


\title{\emph{Ab Initio} Quantum Monte Carlo Simulations of the Uniform Electron Gas \\ without Fixed Nodes II: Unpolarized Case}

\author{T.~Dornheim$^{1,\dagger}$} 
\author{S.~Groth$^{1,\dagger}$} 
\author{T.~Schoof$^1$}
\author{C.~Hann$^{1,2}$}
\author{M.~Bonitz$^1$}\email{bonitz@physik.uni-kiel.de}
\affiliation{$^\dagger$These authors contributed equally to this work.\\ $^1$Institut f\"ur Theoretische Physik und Astrophysik, Christian-Albrechts-Universit\"{a}t zu Kiel, D-24098 Kiel, Germany\\ $^2$Department of Physics, Duke University, Durham, North Carolina 27708, USA}


\date{\today}

\begin{abstract}
In a recent publication [S. Groth \textit{et al.}, PRB (2016)], we have shown that the combination of two novel complementary quantum Monte Carlo approaches, namely configuration path integral Monte Carlo (CPIMC) [T. Schoof \textit{et al.}, PRL
\textbf{115}, 130402 (2015)] and permutation blocking path integral Monte Carlo (PB-PIMC) [T. Dornheim \textit{et al.}, NJP \textbf{17}, 073017 (2015)], allows for the accurate computation of thermodynamic properties of the spin-polarized uniform electron gas (UEG) over a wide range of temperatures and densities without the fixed-node approximation. In the present work, we extend this concept to the unpolarized case, which requires non-trivial enhancements that we describe in detail. We compare our new simulation results with recent restricted path integral Monte Carlo data [E. Brown \textit{et al}., PRL \textbf{110}, 146405 (2013)] for different energy contributions and pair distribution functions and find, for the exchange correlation energy, overall better agreement than for the spin-polarized case, while the separate kinetic and potential contributions substantially deviate.
\end{abstract}

\maketitle
\section{Introduction}
Quantum Monte Carlo (QMC) simulations of fermions are of paramount importantance to describe manifold aspects of nature.
In particular, recent experimental progress with highly compressed matter \cite{fletcher,kraus,regan} such as plasmas in laser fusion experiments \cite{nora,lindl,hu,hurricane,gomez,schmit} and 
solids after laser irradiation \cite{ernst}, but also the need for an appropriate description of compact stars and planet cores \cite{knudson,militzer,nettelmann},
has lead to a high demand for accurate simulations of electrons in the warm dense matter (WDM) regime. Unfortunately, the application of all QMC methods to fermions is severely hampered by the fermion sign problem (FSP) \cite{loh,troyer}.
A popular approach to circumvent this issue is the restricted path integral Monte Carlo (RPIMC) \cite{node} method, which, however, is afflicted with an uncontrollable error due the fixed node approximation\cite{hyd1,hyd2,vfil1,vfil2}.
Therefore, until recently, the quality of the only available QMC results for the uniform electron gas (UEG) in the WDM regime \cite{brown} has remained unclear.

To address this issue, in a recent publication~(paper I, Ref.~\cite{groth}) we have combined two novel complementary approaches: our configuration path integral Monte Carlo (CPIMC) method~\cite{tim1,tim2,prl} excels at high to medium density and arbitrary temperature, while our permutation blocking path integral Monte Carlo (PB-PIMC) approach~\cite{dornheim,dornheim2} significantly extends standard fermionic PIMC \cite{pimc,cep} towards lower temperature and higher density. Surprisingly, it has been found that existing RPIMC results are inaccurate even at high temperatures.

However, although the spin-polarized systems that have been investigated in our previous works are of relevance for the description of e.g.~ferromagnetic materials or strongly magnetized systems, they constitute a rather special case, since most naturally occuring plasmas are predominantly unpolarized.
Therefore, in the present work we modify both our implementations of PB-PIMC and CPIMC to simulate the unpolarized UEG.
So far only a single data set for a small system ($N=14$ electrons, one isotherm) could be obtained in our previous work \cite{prl} because the paramagnetic case turns out to be substantially more difficult than the ferromagnetic one. Therefore, we have developed  novel nontrivial enhancements of our CPIMC algorithm that are discussed in detail.
With these improvements, we are able to present accurate results for different energies for the commonly used case of $N=66$ unpolarized electrons over a broad range of parameters.

Since many details of our approach have been presented in our paper I \cite{groth}, in the remainder of this paper we restrict ourselves to a brief, but selfcontained introduction to CPIMC and PB-PIMC and focus on the differences arising from their application to the unpolarized UEG, compared to the polarized case.
%
In section \ref{jel}, we introduce the model Hamiltonian, both in coordinate space (\ref{cor}) and second quantization (\ref{sec:second_quant}) and, subsequently, provide a brief introduction to the employed QMC approaches (Sec.~\ref{QMC_sec}), namely PB-PIMC (\ref{sec:pb-pimc}) and CPIMC (\ref{sec:CPIMC}).
Finally, in Sec.~\ref{combined}, we present combined results from both methods for the exchange correlation, kinetic, and potential energy (\ref{combined_exc}) as well as the pair distribution function (\ref{combined_pair}). Further, we compare our data to those from RPIMC~\cite{brown}, where available. While we find better agreement than for the spin-polarized case \cite{dornheim2,groth}, there nevertheless appear significant deviations towards lower temperature.

\section{Hamiltonian of the uniform electron gas\label{jel}}

The uniform electron gas (``Jellium'') is a model system of Coulomb interacting electrons in a neutralizing homogeneous background.
As such, it explicitly allows one to study effects due to the correlation and exchange of the electrons, whereas those due to the positive ions are neglected.
Furthermore, the widespread density functional theory (DFT) crucially depends on ab initio results for the exchange correlation energy of the uniform electron gas (UEG),
hitherto at zero temperature \cite{alder}. However, it is widely agreed that the appropriate treatment of matter under extreme conditions requires
to go beyond ground state DFT, which, in turn, needs accurate results for the finite temperature UEG.
While the electron gas itself is defined as an infinite macroscopic system, QMC simulations are possible only for a finite number of particles $N$.
Hence, we always assume periodic boundary conditions and include the interaction of the $N$ electrons in the main simulation cell with all their images via Ewald summation and defer any additional finite-size corrections~\cite{fraser,drummond,lin} to a future publication. 

\subsection{Coordinate representation of the Hamiltonian\label{cor}}
Following Refs.~\cite{fraser,dornheim2}, we express the Hamiltonian (we measure energies in Rydberg and distances in units of the Bohr radius $a_0$) for $N=N_\uparrow+N_\downarrow$ unpolarized electrons in coordinate space as 
 \begin{eqnarray}
  \hat{H} = - \sum_{i=1}^N \nabla^2_i +  \sum_{i=1}^N\sum_{j\ne i}^N e^2 \Psi( \mathbf{r}_i, \mathbf{r}_j) + {N e^2}\xi \; ,
  \label{Hcoord}
 \end{eqnarray}
with the well-known Madelung constant $\xi$ and the periodic Ewald pair interaction
\begin{eqnarray}
\Psi(\mathbf{r}, \mathbf{s} ) &=& \frac{1}{V} \sum_{ \mathbf{G} \ne 0 } \frac{ e^{-\pi^2\mathbf{G}^2/\kappa^2} e^{2\pi i \mathbf{G}(\mathbf{r}-\mathbf{s})} }{ \pi\mathbf{G}^2}
  \label{pair} \nonumber \\ &-& \frac{\pi}{\kappa^2 V} + \sum_\mathbf{R} \frac{ \textnormal{erfc}( \kappa | \mathbf{r}-\mathbf{s} + \mathbf{R} | ) }{ |\mathbf{r}-\mathbf{s}+\mathbf{R} | } \ .
\end{eqnarray}
Here $\mathbf{R}=\mathbf{n}_1L$ and $\mathbf{G}=\mathbf{n}_2/L$ denote the real and reciprocal space lattice vectors, respectively, with the box length $L$, volume $V=L^3$
and the usual Ewald parameter $\kappa$.
Furthermore, PB-PIMC simulations require the evaluation of all forces within the system, where the force between two electrons $i$ and $j$ is given by
\begin{eqnarray}
\label{eq:force}
 \mathbf{F}_{ij} &=&  \frac{2}{V} \sum_{\mathbf{G}\ne 0}\left( \frac{ \mathbf{G} }{ \mathbf{G}^2 }  \textnormal{sin}\left[ 2\pi\mathbf{G}(\mathbf{r}_i - \mathbf{r}_j)\right]e^{- {\pi^2\mathbf{G}^2}/{\kappa^2}} \right)   \\ &+&
 \sum_{\mathbf{R}} \frac{ \mathbf{r}_i - \mathbf{r}_j + \mathbf{R} }{ \alpha^3} \left( \textnormal{erfc}(\kappa\alpha) + \frac{2\kappa\alpha}{\sqrt{\pi}}e^{-\kappa^2\alpha^2} \right)
\; , \nonumber
 \end{eqnarray}
 with the definition $\alpha = | \mathbf{r}_i - \mathbf{r}_j + \mathbf{R} |$.

\subsection{Hamiltonian in second quantization\label{sec:second_quant}}
In second quantization with respect to spin-orbitals of plane waves,
$\langle \mathbf{r} \sigma \;|\mathbf{k}_i\sigma_i\rangle = \frac{1}{L^{3/2}} e^{i\mathbf{k}_i \cdot \mathbf{r}}\delta_{\sigma,\sigma_i}$ with $\mathbf{k}_i=\frac{2\pi}{L}\mathbf{m}_i$, $\mathbf{m}_i\in \mathbb{Z}^3$ and $\sigma_i\in\{\uparrow,\downarrow\}$, 
the model Hamiltonian, Eq.~(\ref{Hcoord}), takes the form
\begin{align}\label{eq:h} 
& \op{H} =
\sum_{i}\mathbf{k}_i^2 \creationop_{i}\annihilop_{i} + 2\smashoperator{\sum_{\substack{i<j,k<l \\ i\neq k,j\neq l}}} 
w^-_{ijkl}\creationop_{i}\creationop_{j} \annihilop_{l} \annihilop_{k} + Ne^2\xi,
\end{align}
with the antisymmetrized two-electron integrals, $w^-_{ijkl} =w_{ijkl}-w_{ijlk}$, where
\begin{align} 
\; w_{ijkl}=\frac{4\pi e^2}{L^3 (\mathbf{k}_{i} - \mathbf{k}_{k})^2}\delta_{\mathbf{k}_i+\mathbf{k}_j, \mathbf{k}_k + \mathbf{k}_l}\delta_{\sigma_i,\sigma_k}\delta_{\sigma_j,\sigma_l}\;,
\label{eq:two_ints}
\end{align}
and the Kronecker deltas ensuring both momentum and spin conservation. The first (second) term in the Hamiltonian Eq.~(\ref{eq:h}) describes the kinetic (interaction) energy.
The operator 
$\creationop_{i}$  ($ \annihilop_{i}$) 
creates (annihilates) a particle in the spin-orbital $|\mathbf{k}_i\sigma_i\rangle$.

\section{Fermionic quantum Monte Carlo\label{QMC_sec} without fixed nodes}
Throughout the entire work, we consider the canonical ensemble, i.e., the volume $V$, particle number $N$ and inverse temperature $\beta=1/k_\text{B}T$ are fixed.
In equilibrium statistical mechanics, all thermodynamic quantities can be derived from the partition function
\begin{eqnarray}
 Z = \text{Tr} \op{\rho} \; , \label{eq:partition}
\end{eqnarray}
which is of central importance for any QMC formulation and defined as the trace over the canonical density operator
\begin{eqnarray}
 \hat{\rho} = e^{-\beta\op{H}} \; .
\end{eqnarray}
The expectation value of an arbitrary operator $\op{A}$ is given by
\begin{eqnarray}
 \braket{\op{A}} = \frac{ \text{Tr} ( \op{A}\op{\rho} ) }{ \text{Tr}\op{\rho} } = \frac{1}{Z} \text{Tr}(\op{A}\op{\rho}) \; . \label{eq:expectation}
\end{eqnarray}
However, for an appropriate description of fermions, Eqs.~(\ref{eq:partition}) and (\ref{eq:expectation}) must be extended either by antisymmetrizing $\op{\rho}\to\op{\rho}^-$ 
or the trace itself~\cite{tim1}, $\text{Tr}\to\text{Tr}^-$. Therefore, it holds
\begin{eqnarray}
 Z = \text{Tr} \op{\rho}^- = \text{Tr}^- \op{\rho} \; . \label{eq:p2}
\end{eqnarray}
While defining the trace in Eq.~(\ref{eq:p2}) as either expression does not change the well-defined thermodynamic expectation values,
it does lead to rather different formulations of the same problem.
The combination of antisymmetrizing the density matrix and evaluating the trace in coordinate space is the first step towards both standard PIMC and PB-PIMC, cf.~Sec.~\ref{sec:pb-pimc}, but also RPIMC.
All these approaches share the fact that they are efficient when fermionic quantum exchange does not yet dominate a systm, but they
will become increasingly costly towards low temperature and high density.
Switching to second quantization and carrying out the trace in antisymmetrized Fock space, on the other hand, is the basic idea behind our CPIMC method, cf.~Sec.~\ref{sec:CPIMC}, and, in a different way, behind the likewise novel density matrix QMC method \cite{blunt}. The latter approach has recently been applied to the the case of $N=4$ spin-polarized electrons~\cite{malone}, where complete agreement with our CPIMC results~\cite{tim2} was reported. 
These QMC approaches tend to excel at high density, i.e., weak nonideality, and become eventually unfeasible towards stronger coupling strength.

Therefore, it is a natural strategy to combine different representations at complementary parameter ranges as this does effectively allow to circumvent the numerical shortcomings with which  
every single fermionic QMC method is necessarily afflicted \cite{dornheim2,groth}.

\subsection{Permutation blocking PIMC\label{sec:pb-pimc}}

\subsubsection{Basic idea}
In this section, we will briefly introduce our permutation blocking PIMC approach.
A more detailed description of the method itself and its application to the spin-polarized UEG can be found in Refs.~\cite{dornheim,dornheim2}.

The basic idea behind PB-PIMC is essentially equal to standard PIMC in coordinate space, e.g., Ref.~\cite{cep}, but, in addition,
combines two well-known concepts: 1) antisymmetric imaginary time propagators, i.e., determinants \cite{det1,det2,det3}, and 2) a fourth-order factorization of the density matrix \cite{ho1,ho2,ho3,ho4}.
Furthermore, since this leads to a significantly more complicated configuration space without any fixed paths, one of us has developed an efficient set of Metropolis Monte Carlo~\cite{metropolis} updates
that utilize the temporary construction of artificial trajectories~\cite{dornheim}. 
As mentioned above, we evaluate the trace within the canonical partition function for $N=N_\uparrow+N_\downarrow$ unpolarized electrons in coordinate representation
\begin{eqnarray}
 Z &=& \frac{1}{N_\uparrow!N_\downarrow!} \sum_{\sigma_\uparrow\in S_{N_\uparrow}}\sum_{\sigma_\downarrow\in S_{N_\downarrow}} \textnormal{sgn}(\sigma_\uparrow)\ \textnormal{sgn}(\sigma_\downarrow) \nonumber \\ & &
 \times \int \textnormal{d}\mathbf{R}\ \bra{ \mathbf{R} } e^{-\beta\op{H}} \ket{ \op{\pi}_{\sigma_\uparrow}\op{\pi}_{\sigma_\downarrow} \mathbf{R}} \;, \label{boseZ}
\end{eqnarray}
with $\op{\pi}_{\sigma_{\uparrow,\downarrow}}$ being the exchange operator that corresponds to a particular element $\sigma_{\uparrow,\downarrow}$ from the permutation group $S_{N_{\uparrow,\downarrow}}$ with associated sign $\textnormal{sgn}(\sigma_{\uparrow,\downarrow})$
and $\uparrow$ ($\downarrow$) denoting spin-up (spin-down) electrons.
However, since the kinetic and potential contributions to the Hamiltonian, $\op{K}$ and $\op{V}$, do not commute, the low-temperature matrix elements of $\op{\rho}$ are not known.
To overcome this issue, we use the common group property $\op\rho (\beta) = \prod_{\alpha=0}^{P-1} \op\rho(\epsilon)$ of the density matrix, 
with $\epsilon=\beta/P$, and approximate each of the $P$ factors at a $P$ times higher temperature by the fourth-order factorization \cite{ho2,ho3}
\begin{eqnarray}
 \label{cchin} e^{-\epsilon\op{H}} &\approx& e^{-v_1\epsilon\op{W}_{a_1}} e^{-t_1\epsilon\op{K}} e^{-v_2\epsilon\op{W}_{1-2a_1}} \nonumber \\ & & \times e^{-t_1\epsilon\op{K}} e^{-v_1\epsilon\op{W}_{a_1}} e^{-2t_0\epsilon\op{K}}\; .
\end{eqnarray}
It should be noted that Eq.~(\ref{cchin}) allows for sufficient accuracy, even for small $P$.
The $\op{W}$ operators in Eq.\ (\ref{cchin}) denote a modified potential that combines the usual potential energy $\op{V}$ with
double commutator terms of the form 
\begin{eqnarray}
 [[\op{V},\op{K}],\op{V}] = \frac{\hbar^2}{m} \sum_{i=1}^N |\mathbf{F}_i|^2 \;, \; \mathbf{F}_i = -\nabla_i V(\mathbf{R})\; ,
\end{eqnarray}
and, therefore, require the evaluation of all forces within the system, cf.~Eq.~(\ref{eq:force}).
The final result for the PB-PIMC partition function is given by
\begin{eqnarray}
\label{finalz} Z &=&  \nonumber \frac{1}{(N_\uparrow!N_\downarrow!)^{3P}} \int \textnormal{d}\mathbf{X} \\ & &  \prod_{\alpha=0}^{P-1} \Big( e^{-\epsilon\tilde V_\alpha}e^{-\epsilon^3u_0\frac{\hbar^2}{m}\tilde F_\alpha} D_{\alpha,\uparrow}D_{\alpha,\downarrow} \Big) \; ,
\end{eqnarray}
with  $\tilde V_\alpha$ and $\tilde F_\alpha$ containing all contributions of the potential energy and the forces, respectively.
The exchange-diffusion functions are defined as
\begin{eqnarray}
 D_{\alpha,\uparrow} &=& \textnormal{det}(\rho_{\alpha,\uparrow} )\textnormal{det}(\rho_{\alpha A,\uparrow} )\textnormal{det}(\rho_{\alpha B,\uparrow} ) \\ \nonumber
  D_{\alpha,\downarrow} &=& \textnormal{det}(\rho_{\alpha,\downarrow} )\textnormal{det}(\rho_{\alpha A,\downarrow} )\textnormal{det}(\rho_{\alpha B,\downarrow} )
\end{eqnarray}
and contain the determinants of the diffusion matrices
\begin{eqnarray}
 \rho_{\alpha,\uparrow}(i,j)  =    \lambda_{t_1\epsilon}^{-3}\sum_\mathbf{n} e^{ -\frac{\pi}{\lambda^2_{t_1\epsilon}} ( \mathbf{r}_{\alpha,\uparrow,j} - \mathbf{r}_{\alpha A,\uparrow,i}+\mathbf{n}L)^2  }\;, \label{diffusion}
\end{eqnarray}
with $\lambda_{t_1\epsilon}=\sqrt{2\pi\epsilon t_1\hbar^2/m}$ being the thermal
wavelength of a single ``time slice''.

In contrast to standard PIMC, where each permutation cycle has to be explicitly sampled,
we combine both positively and negatively signed configuration weights in the determinants both for the spin-up and spin-down electrons.
This leads to a cancellation of many terms and, consequently, a significantly increased average sign in our Monte Carlo simulations.
Yet, this ``permutation blocking'' is only effective when $\lambda_{t_1\epsilon}$ is comparable to the mean inter-particle distance, i.e.,
when there are both large diagonal and off-diagonal elements in the diffusion matrices.
With an increasing number of high-temperature factors $P$, $\lambda_{t_1\epsilon}$ decreases and, eventually, when there is only but a single
large element in each row of the $\rho_{\alpha,\uparrow}$, the average sign converges towards that of standard PIMC. 
For this reason, it is crucial to combine the determinants from the antisymmetric propagators with a higher order factorization of the density matrix, cf.~Eq.~(\ref{cchin}).
It is only this combination which allows for sufficient accuracy with as few as two or three propagators while, at the same time, the benefit of the blocking within the determinants is maximized.
Furthermore, we note that electrons with different spin-projections do not exchange at all. Therefore, PB-PIMC simulations of the unpolarized UEG with $N=N_\uparrow+N_\downarrow$ do suffer from a significantly less severe sign problem than for $N=2N_\uparrow$ spin-polarized electrons.

\subsubsection{Application to the unpolarized UEG}
  \begin{figure}[]
 \centering
 \includegraphics[width=0.49\textwidth]{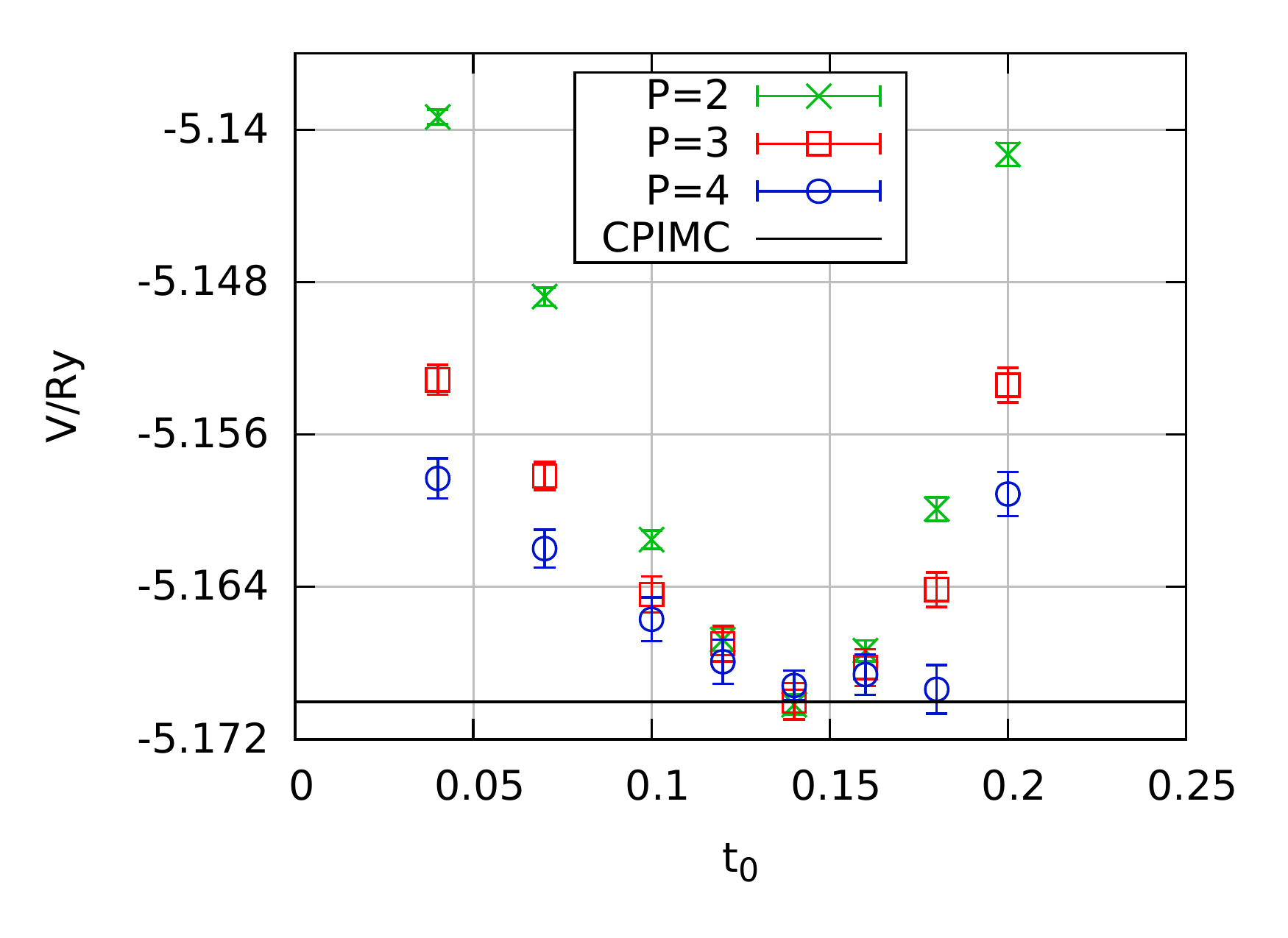}
 \caption{\label{optimum}Influence of the relative interslice spacing $t_0$ on the convergence --
 The potential energy from PB-PIMC simulations of $N=4$ unpolarized electrons at $\theta=0.5$ and $r_s=1$ is plotted versus $t_0$ for the fixed choice $a_1=0.33$.
 }
\end{figure}

The accuracy of our PB-PIMC simulations crucially depends on the systematic error due to the employed higher order factorization \cite{dornheim,dornheim2}. Thus, we begin the investigation of the unpolarized electron gas with the analysis of the convergence behavior with respect to the two free parameters from Eq.~(\ref{cchin}), namely $a_0$ (weighting the contributions of the forces on different time slices) and $t_0$ (controlling the relative interslice spacing).
In Fig.~\ref{optimum}, we set $a_0=0.33$ fixed, which corresponds to equally weighted forces on all slices, and plot the potential energy for $P=2,3,4$ over the entire $t_0$-range for a benchmark system of $N=4$ unpolarized electrons at $r_s=1$ and $\theta=0.5$. Evidently, for all $t_0$ values $V$ converges monotonically from above towards the exact result, which has been obtained with CPIMC. The optimum value for $t_0$ is located around $t_0=0.14$, where all three PB-PIMC values are within single error bars with the black line. For completeness, we mention that this particular set of the optimum free parameters for the energy is consistent with the
previous findings for different systems \cite{ho3,dornheim,dornheim2}.

  \begin{figure}[]
 \centering
 \includegraphics[width=0.49\textwidth]{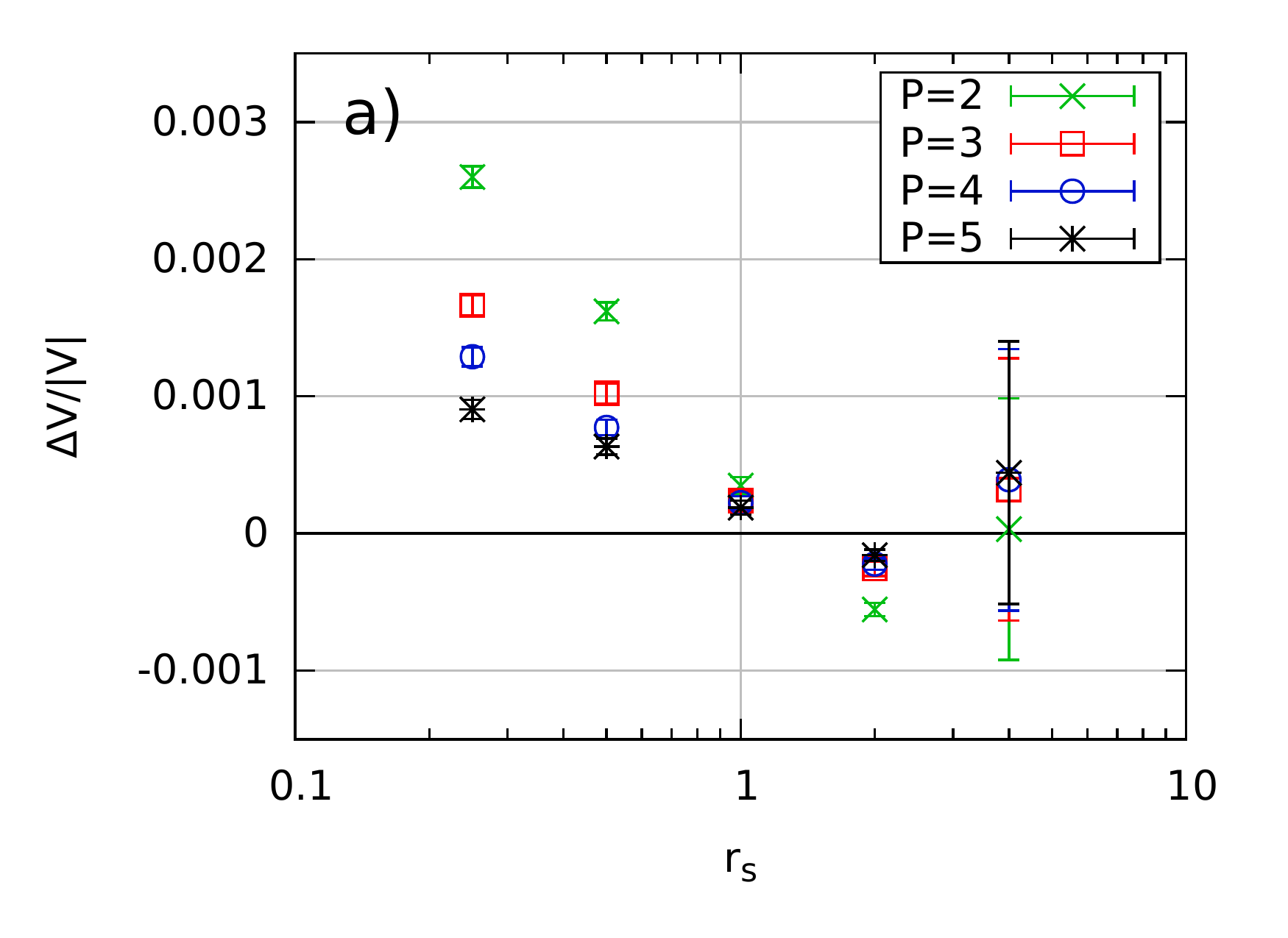}
 \includegraphics[width=0.49\textwidth]{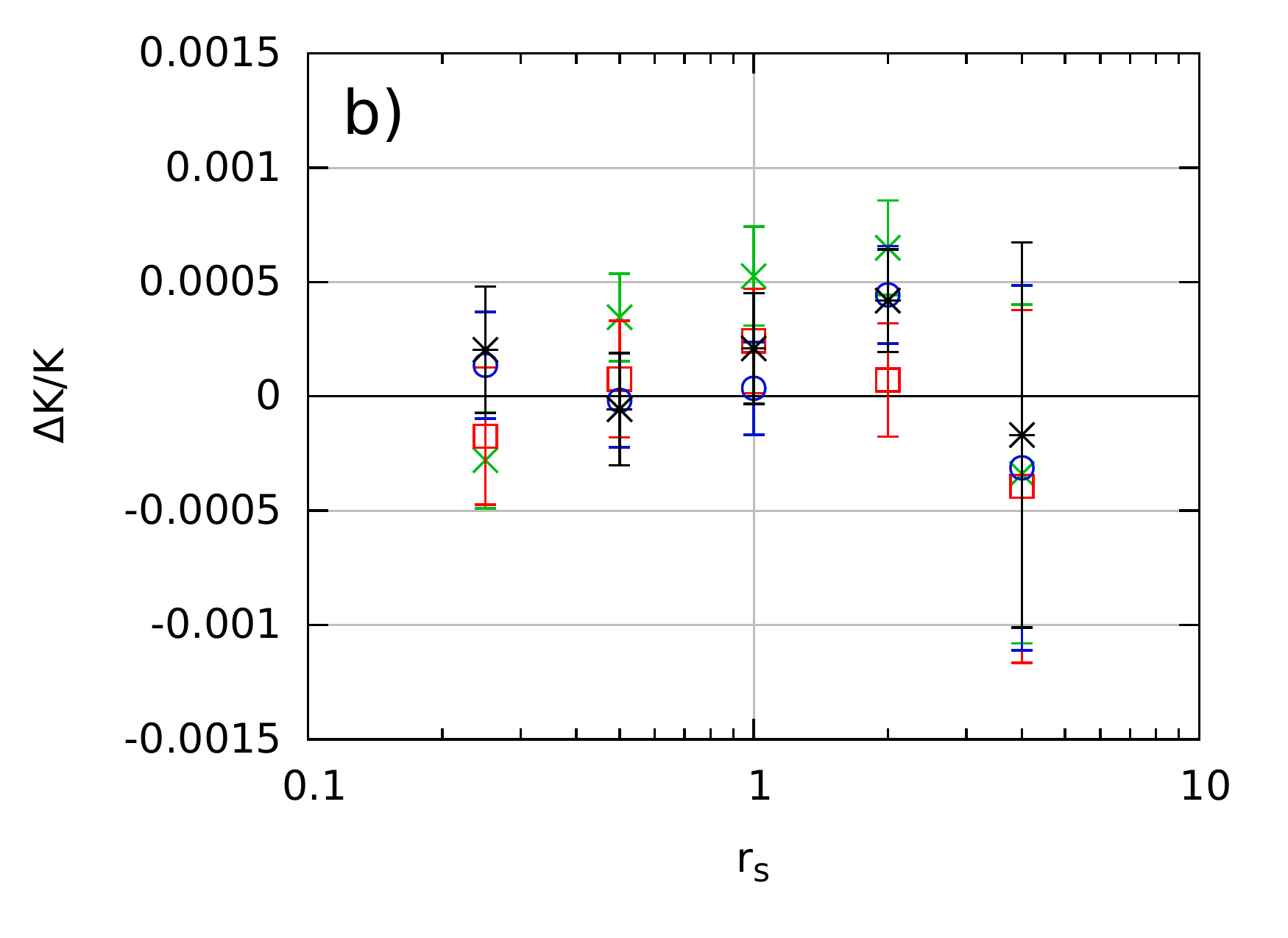}
 \caption{\label{error}Density dependence of the relative time step error from PB-PIMC with $a_1=0.33$ and $t_0=0.14$ --
 The relative differences between PB-PIMC results with $P=2,3,4,5$ and reference data from CPIMC are plotted versus $r_s$ for the potential energy (a) and the kinetic energy (b).
 }
\end{figure}
A natural follow-up question is how the convergence with $P$ behaves with respect to the density parameter $r_s$.
In Fig.~\ref{error}, we show results for the relative error of the potential ($\Delta V/|V|$, panel a) and kinetic energy ($\Delta K/K$, panel b), where the reference values are again obtained from CPIMC. The statistical uncertainty is mainly due to PB-PIMC, except for $r_s=4$ where the CPIMC error bar predominates.
For the kinetic energy, even for $P=3$ there are no clear systematic deviations from the exact result over the entire $r_s$-range. Only with two propagators, our results for $K$ appear to be slightly too large for $r_s\in(0.5,1,2)$, although this trend hardly exceeds $\Delta K/K=5\times 10^{-4}$.
For the potential energy, the factorization error behaves quite differently. For $r_s \ge 1$, even with two propagators the accuracy is better than $0.1\%$, while towards higher density ($r_s < 1$), the convergence significantly deteriorates. In particular, at $r_s=0.25$ even with $P=5$
there is a deviation of $\Delta V/|V|\approx 0.1\%$.
This observation is in striking contrast to our previous investigation of the polarized UEG, where the relative error in both $K$ and $V$ decreased towards $r_s\to0$. 
The reason for this trend lies in the presence of two different particle species which do not exchange with each other, namely $N_\uparrow$ spin-up and $N_\downarrow$ spin-down electrons. Even at high density, two electrons from the same species are effectively separated by their overlapping kinetic density matrices that cancel in the determinants, which is nothing else than the Pauli blocking.
Yet, a spin-up and a spin-down electron do not experience such a repulsion and, at weak coupling (small $r_s$), can be separated by much smaller distances $r$ from each other.
With decreasing $r$ the force terms in Eq.~(\ref{cchin}) that scale as $F(r)\propto 1/r^2$ will eventually exceed the Coulomb potential $V(r)\propto 1/r$, i.e., the higher order correction predominates. This trend must be compensated by an increasing number of propagators $P$.
Hence, the fermionic nature of the electrons that manifests as the Pauli blocking significantly enhances the performance of our factorization scheme, which means that the simulation of unpolarized systems is increasingly hampered towards high density. In addition to the Monte Carlo inherent sign problem, this is a further reason to combine PB-PIMC with CPIMC, since the latter excels just in this regime.

\begin{figure}[]
 \centering
 \includegraphics[width=0.49\textwidth]{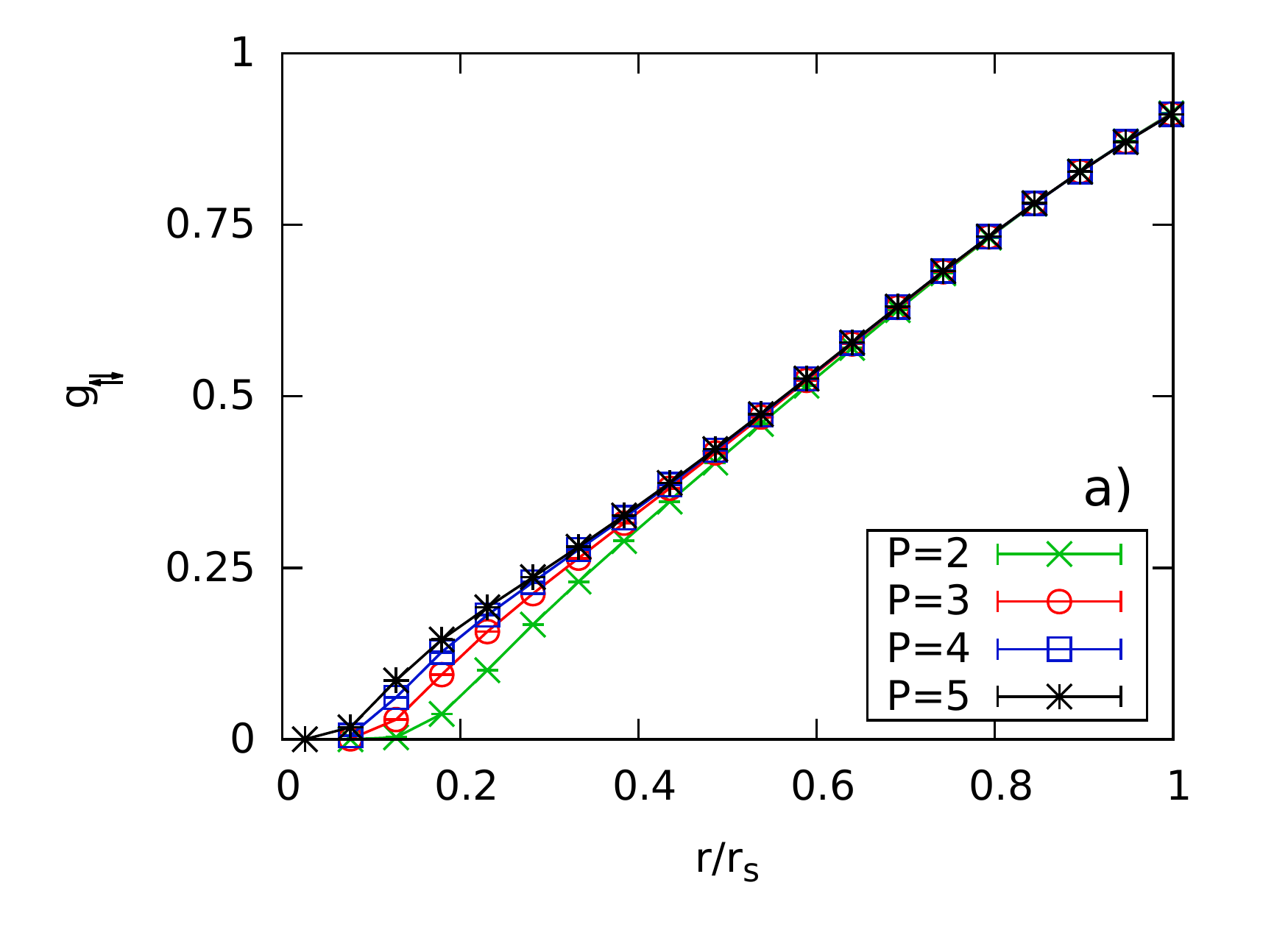}
 \includegraphics[width=0.49\textwidth]{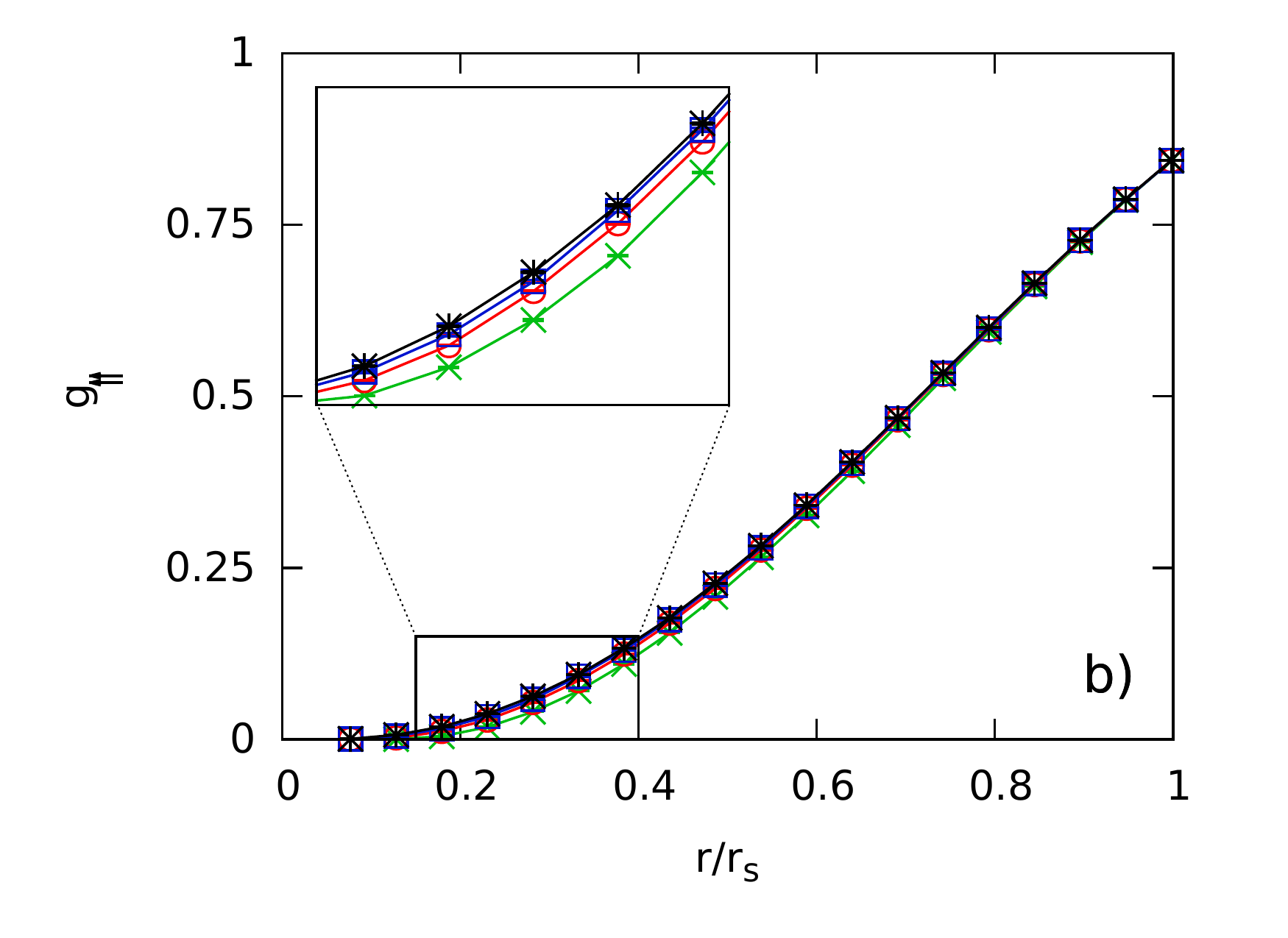}
 \caption{\label{cf_e}Convergence of the pair distribution function for $N=4$ unpolarized electrons at $\theta=1$ and $r_s=4$ --
 Shown are PB-PIMC results for the inter- ($g_{\uparrow\downarrow}$, panel a) and intra-species ($g_{\uparrow\uparrow}$, panel b) distribution
 function for different numbers of propagators $P$ and the fixed free parameters $a_1=0.33$ and $t_0=0.14$.
 }
\end{figure}

In our recent analysis of PB-PIMC for electrons in a $2D$ harmonic trap \cite{dornheim}, it was found that, while the combination $a_0=0.33$ and $t_0=0.14$ (parameter set a) is favorable for a fast convergence of the energy, it does not perform so well for other properties like, in that case, the density profile.
To address this issue, we again simulate a benchmark system of $N=4$ unpolarized electrons
and compute the pair distribution function $g(r)$, see, e.g.~Ref.~\cite{gori} for a comprehensive discussion. In Fig.~\ref{cf_e}, we show results for the above combination of free parameters (a) and $P=2,3,4,5$. Panel a) displays the data for the inter-species distribution function $g_{\uparrow\downarrow}$. We note that, for the infinite UEG, this quantity approaches unity at large distances, but the small simulation box for $N=4$ restricts us to the depicted $r$-range.
All four curves deviate from each other for $r\lesssim0.5$, which indicates that $g_{\uparrow\downarrow}$ is not yet converged even for $P=5$ at small distances, and are equal otherwise. This is again a clear indication of the shortcomings of our fourth-order factorization, which overestimates the Coulomb repulsion at short ranges.
The intra-species distribution function $g_{\uparrow\uparrow}=g_{\downarrow\downarrow}$, which is shown in panel b), does not exhibit such a clear trend since only the green curve that corresponds to $P=2$ can be distinguished from the rest. This is, of course, expected and a consequence of the Pauli blocking as explained above. 

\begin{figure}[]
 \centering
 \includegraphics[width=0.49\textwidth]{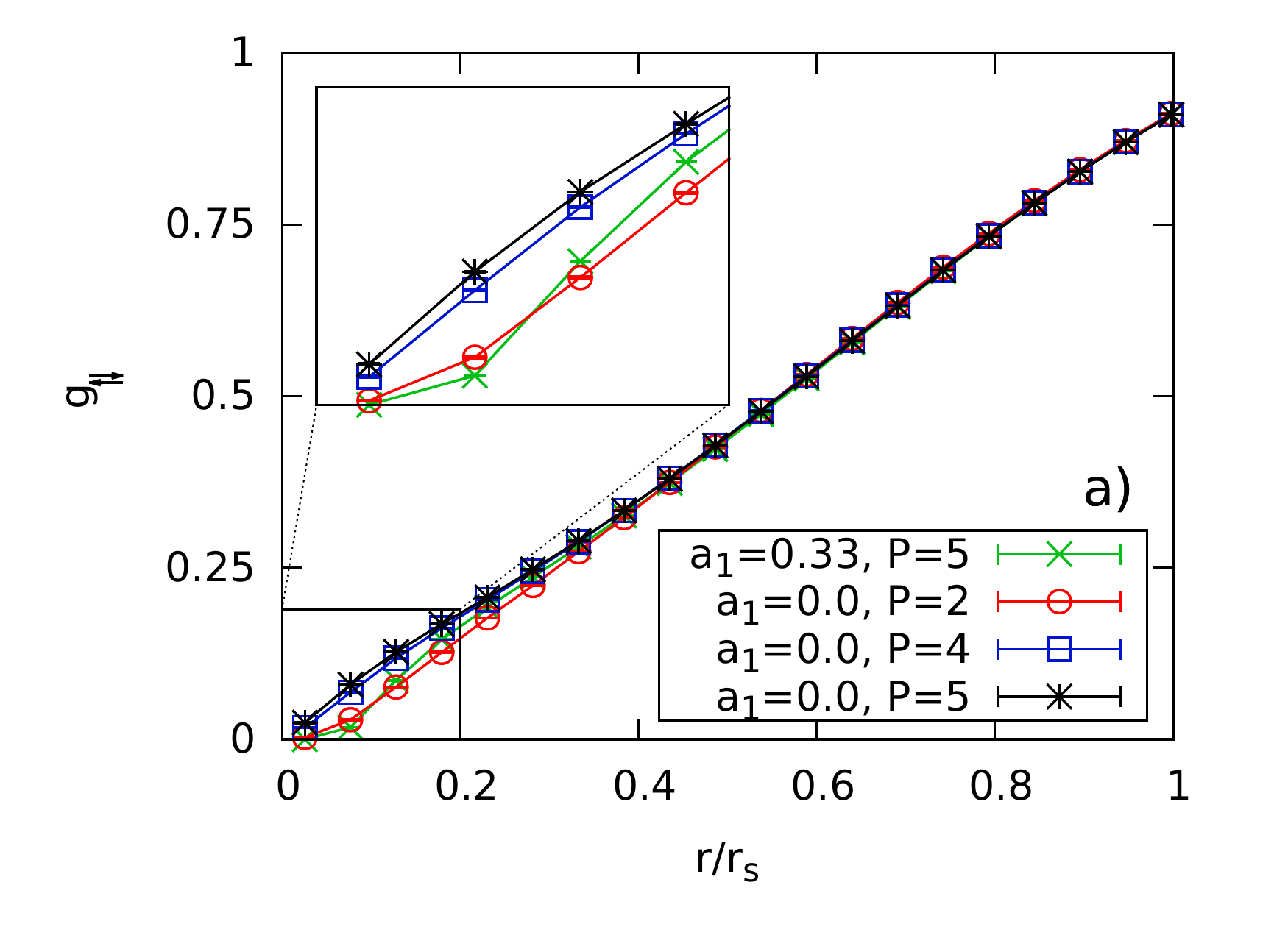}
 \includegraphics[width=0.49\textwidth]{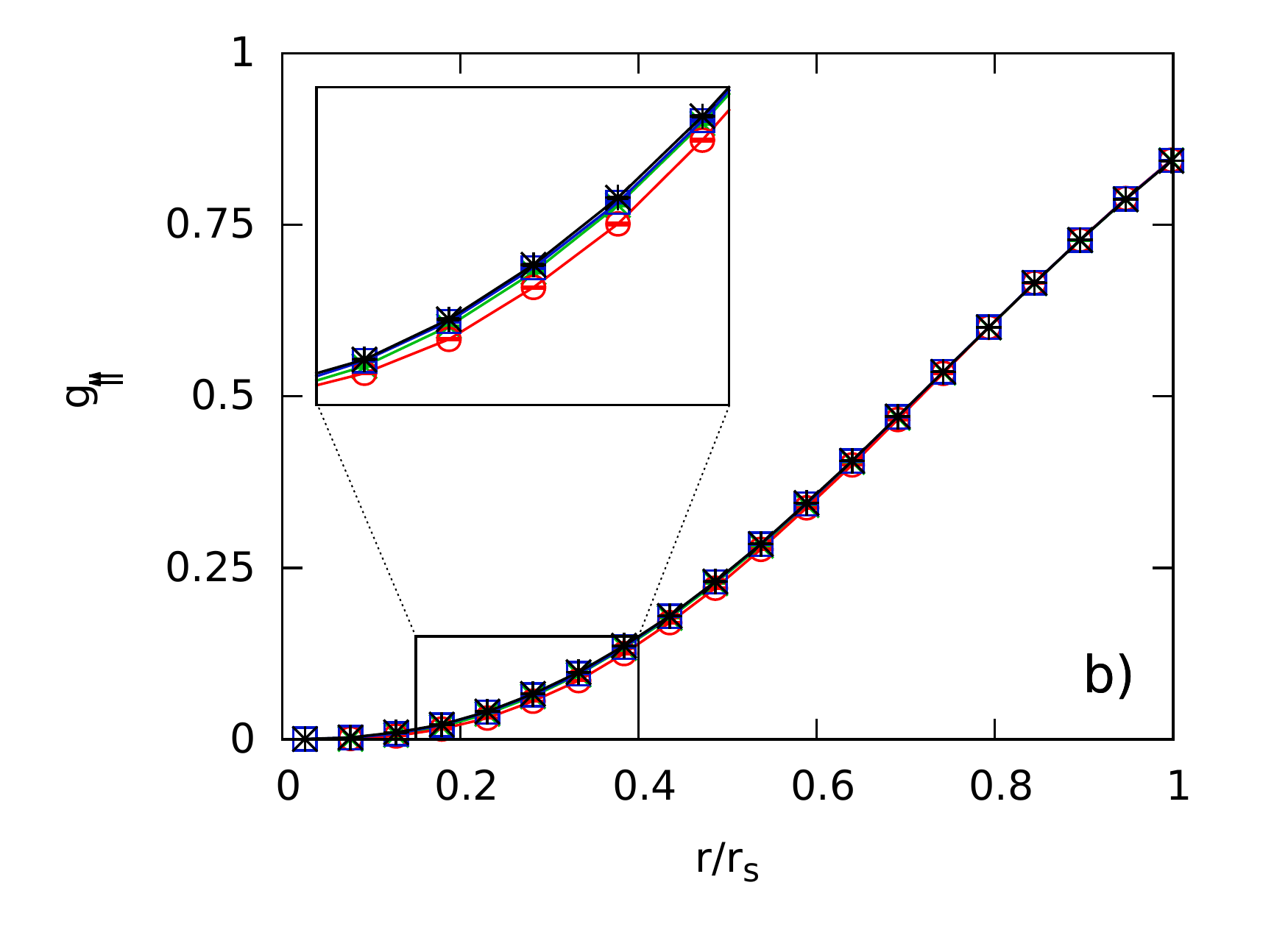}
 \caption{\label{cf_c}onvergence of the pair distribution function for $N=4$ unpolarized electrons at $\theta=1$ and $r_s=4$  --
 Shown is the same information as in Fig.~\ref{cf_e}, but for a different combination of free parameters, i.e., $a_0=0$ and $t_0=0.04$.
  }
\end{figure}

Evidently, our propagator with the employed choice of free parameters (a) does not allow for an accurate description of the Coulomb repulsion at short distances. 
To understand this issue, we repeat the simulations with a different combination $a_0=0$ and $t_0=0.04$ (parameter set b), which has already proven to be superior to parameter set (a) for the radial density in the $2D$ harmonic trap.
The results are shown in Fig.~\ref{cf_c} for different numbers of propagators. The data with $P=2$  are nearly equal to the results from parameters (a) and $P=5$. 
The data for $P=4$ and $P=5$ almost coincide and are significantly increased with respect to the other curves. The main reason for the improved convergence of parameter set (b) is the choice $a_0=0$, which means that the forces are only taken into account on intermediate time slices. Due to the diagonality of the pair distribution function in coordinate space, it is measured exclusively on the main slices, for whose distribution the force terms do not directly enter. For this reason, the inter-species pair distribution function is not as drastically affected by the divergence of the $F(r)\propto1/r^2$ terms at small $r$ and the convergence of this quantity is significantly improved.
For completeness, in panel b) we again show results for $g_{\uparrow\uparrow}$, which, for parameter set (b), are almost converged even for two propagators.
It is important to note that while the description of the Coulomb repulsion at very short ranges is particularly challenging, this does not predominate in larger systems since the average number of particles within distance $r\in[\tilde r,\tilde r + \Delta\tilde r)$ increases as $N(\tilde r)\propto \tilde r^2$.
For $N=66$ unpolarized electrons, which is the standard system size within this work, these effects are by far not as important and, for the same combination of $r_s$ and $\theta$ as in Fig.~\ref{cf_c}, both the inter- and intra-species distribution function are of much higher quality, cf.~Fig.~\ref{brown_cf}.
  \begin{figure}[]
 \centering
 \includegraphics[width=0.49\textwidth]{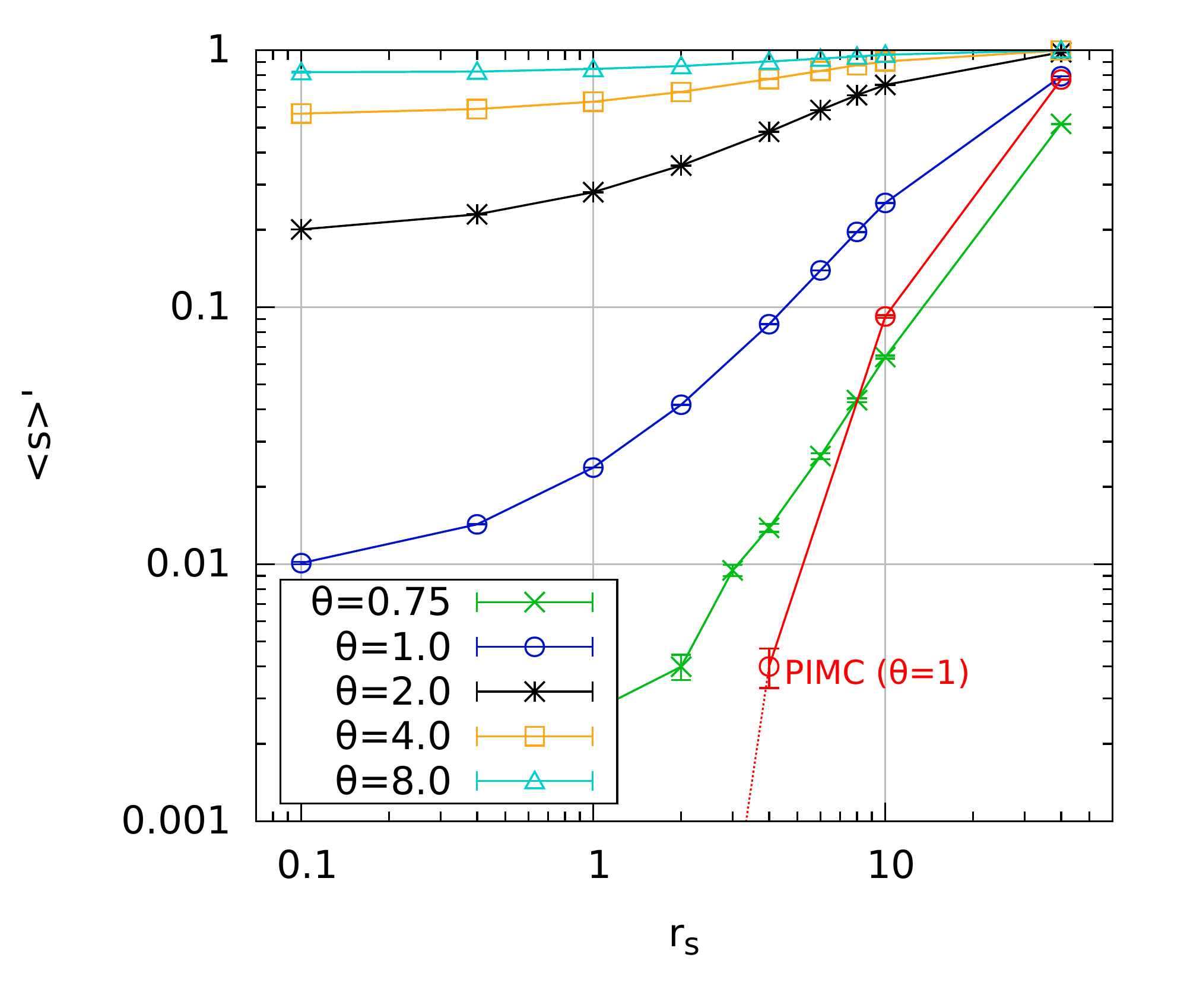}
 \caption{\label{pb_sign}Average sign for PB-PIMC simulations of $N=66$ unpolarized electrons at different temperatures -- All PB-PIMC data have been obtained for $P=2$ with $a_0=0.33$ and $t_0=0.14$ and the standard PIMC data (red curve) have been taken from Ref.~\cite{brown}.
 }
\end{figure}

Up to this point, only data for small benchmark systems with $N=4$ electrons have been presented. To obtain meaningful results for the UEG, we simulate $N=66$ unpolarized electrons, which is a commonly used model system since it corresponds to a closed momentum shell and, therefore, is well suited as a starting point for an extrapolation to the thermodynamic limit (finite size corrections).
In Fig.~\ref{pb_sign}, the average sign, cf.~Eq.~(\ref{eq:average}), is plotted versus the density parameter $r_s$ for five different temperatures. For $\theta=2,4,8$, $\braket{s}'$ is almost equal to unity for $r_s=40$ and decreases just a trifle towards higher density, until it saturates at $r_s\sim0.5$. 
Consequently, simulations are possible over the entire density range with relatively small computational effort. The slight increase of $\braket{s}'$ around $r_s\in[1,10]$ is a nonideality effect: At high density, the system is approximately ideal and the Fermi temperature $\theta_\textnormal{F}$ is an appropriate measure for quantum degeneracy. With increasing $r_s$, coupling effects become more important, which leads to a stronger separation of the electrons. Thus, there is less overlap of the kinetic density matrices and the determinants become exclusively positive.
For $\theta=1$, the average sign already significantly deviates from unity at $r_s=40$ and exhibits a more severe decrease towards smaller $r_s$. Nevertheless, it attains a finite value $\braket{s}'\approx 0.01$ 
even at high density $r_s=0.1$, which means that simulations are more involved but still manageable over the entire coupling range. This is in stark contrast to standard PIMC without the permutation blocking (red circles), for which the sign exhibits a sharp drop and simulations become unfeasible below $r_s\approx5$.
Finally, the green curve corresponds to $\theta=0.75$ where PB-PIMC is capable to provide accurate results for $r_s\ge3$.

\subsection{Configuration PIMC\label{sec:CPIMC}}

\subsubsection{Basic idea}
In this section, the main aspects of our CPIMC approach are explained.
A detailed derivation of the CPIMC expansion of the partition function
and the utilized Monte Carlo steps for the polarized UEG can be found in Refs.~\cite{tim2,groth}.

\begin{figure}
\includegraphics[width=85mm]{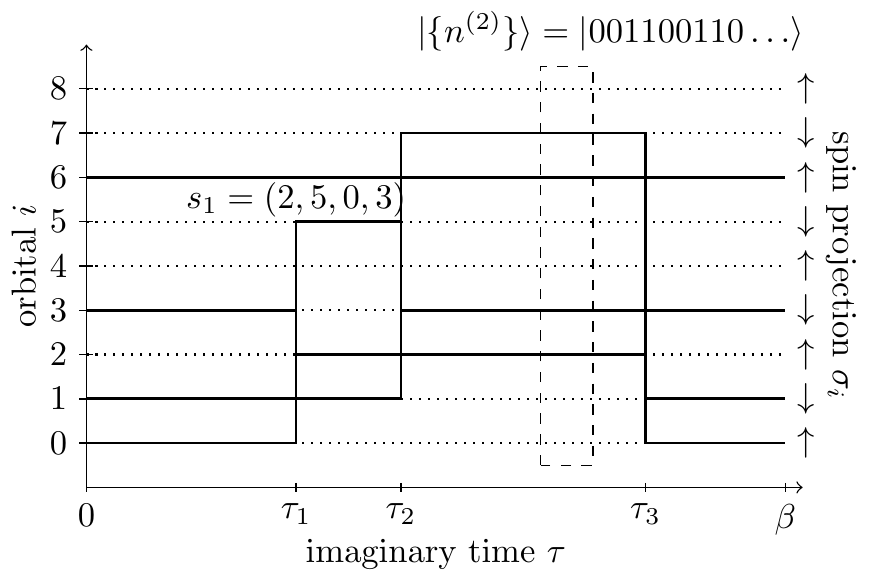}
 \caption{Typical closed path of $N=4$ unpolarized particles in Slater determinant (Fock) space. The state with four occupied orbitals $|\mathbf{k}_0\uparrow\rangle, |\mathbf {k}_1\downarrow\rangle, |\mathbf {k}_3\downarrow\rangle, |\mathbf {k}_6\uparrow\rangle$ undergoes a two-particle excitation $s_1$ at time $\tau_1$ replacing the occupied orbitals $\ket{\mathbf{k}_0\uparrow}, |\mathbf {k}_3\downarrow\rangle$ by 
$|\mathbf{k}_2\uparrow\rangle, |\mathbf{k}_5\downarrow\rangle$. Two further excitations occur at $\tau_2$ and $\tau_3$.
The states at the ``imaginary times'' $\tau = 0$ and $\tau = \beta$ coincide. In addition, the total spin projection is conserved at any time. All possible paths contribute to the partition function $Z$, Eq.~(\ref{eq:Z_expansion}).}
 \label{fig:sketch}
\end{figure}

For CPIMC, instead of evaluating the trace of the partition function Eq.~(\ref{eq:partition}) in coordinate representation, we switch to second quantization and perform the trace with anti-symmetrized $N-$particle states (Slater-determinants)
\begin{align}
|\{n\}\rangle=|n_1, n_2, \dots\rangle\;,
\end{align}
with $n_i$ being the fermionic occupation number ($n_i\in\{0, 1\}$) of the $i$-th spin-orbital $|\mathbf{k}_i\sigma_i\rangle$, where we choose the ordering of orbitals such that even (odd) orbital numbers have spin-up (spin-down) $\sigma=\uparrow(\downarrow)$. In this representation, fermionic anti-symmetry is automatically taken into account via the anti-commutation relations of the creation and annihilation operators, and thus, an explicit anti-symmetrization of the density operator is not needed. The expansion of the partition function is based on the concept of continuous time QMC, e.g., Refs.~\cite{prokofiev96,prokofiev98}, where the Hamiltonian is split into a diagonal and off-diagonal part $\op{H}=\op{D}+\op{Y}$ with respect to the chosen basis. Summing up the entire perturbation series of the density operator $e^{-\beta\op{H}}$ in terms of $\op{Y}$ finally yields
\begin{align}
Z = &
\sum_{K=0,\atop K \neq 1}^{\infty} \sum_{\{n\}}
\sum_{s_1\ldots s_{K-1}}\,
\int\limits_{0}^{\beta} d\tau_1 \int\limits_{\tau_1}^{\beta} d\tau_2 \ldots \int\limits_{\tau_{K-1}}^\beta d\tau_K 
\label{eq:Z_expansion} \\\nonumber
& (-1)^K  
e^{-\sum\limits_{i=0}^{K} D_{\{n^{(i)}\}} \left(\tau_{i+1}-\tau_i\right) } 
\prod_{i=1}^{K} Y_{\{n^{(i)}\},\{n^{(i-1)}\} }(s_i)\;,
\end{align}
with the Fock space matrix elements of the diagonal and off-diagonal operator
\begin{align}
 &D_{\{n^{(i)}\}} = \sum_l \mathbf{k}_l^2 n^{(i)}_{l} + \sum_{l<k}w^-_{lklk}n^{(i)}_{l}n^{(i)}_{k} \;,\label{eq:diagonal}\\
&Y_{ \{n^{(i)}\},\{n^{(i-1)}\} }(s_i) =w^-_{s_i}(-1)^{\alpha^{\phantom{-}}_{s_i}}\;.
\label{eq:off_diagonal}
\end{align}
Here, $s_i=(pqrs)$ defines the four occupation numbers in which  $\{n^{(i)}\}$ and $\{n^{(i-1)}\}$ differ, where it is $p<q$ and $r<s$. In this notation, the exponent of the fermionic phase factor is given by  
\begin{align}
\alpha^{\phantom{-}}_{s_i} &=\alpha^{(i)}_{pqrs}=\sum_{l=p}^{q-1}n^{(i-1)}_{l}+\sum_{l=r}^{s-1}n^{(i)}_{l}\;.
\nonumber
\end{align}
Due to the trace, each summand in Eq.~(\ref{eq:Z_expansion}) fulfills $\{n\}=\{n^{(0)}\}=\{n^{(K)}\}$ and hence can be interpreted as a $\beta$-periodic path in Fock space. An example of such a path for the case of an unpolarized UEG is depicted in Fig.~\ref{fig:sketch}. 
The starting determinant $\{n\}$ at $\tau=0$ undergoes $K$ excitations of type $s_i$ at time $\tau_i$, which we refer to as "kinks". The weight of each path is computed according to the second line of Eq.~(\ref{eq:Z_expansion}), which can be both positive and negative. Since the Metropolis algorithm\cite{metropolis} can only be applied to strictly positive weights, we have to take the modulus of the weights in our MC procedure and compute expectation values according to
\begin{align}
\langle O \rangle = \frac{\langle Os\rangle^\prime}{\langle s \rangle^\prime}\;, 
\label{eq:average}
\end{align}
where $O$ is the corresponding Monte Carlo estimator of the observable, $\langle\cdot\rangle^\prime$ denotes the expectation value with respect to the modulus weights, and $s$ measures the sign of each path. Therefore, $\braket{s}^\prime$ is the \emph{average sign} of all sampled paths during the MC simulation. It is straightforward to show that the relative statistical error of observables computed according to Eq.~(\ref{eq:average}) is inversely proportional to the average sign. As a consequence, in practice, reliable expectation values can be obtained if the average sign is larger than about $10^{-4}$. 

\subsubsection{Application to the unpolarized UEG}
\begin{figure}[]
 \centering
 \includegraphics{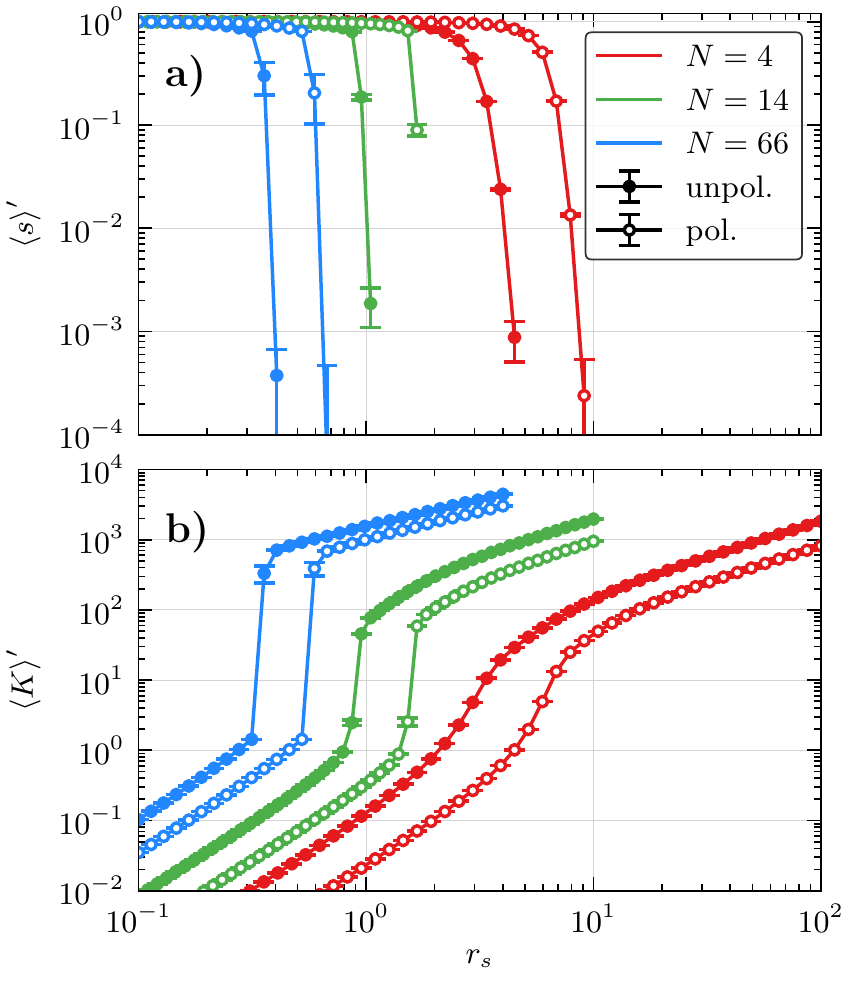}
 \caption{Average sign \textbf{a)} and average number of kinks \textbf{b)} of direct CPIMC, plotted versus the density parameter for three different particle numbers $N=4,14,66$ in $N_B=2109,4169,5575$ plane wave basis functions, respectively, at $\theta=1$. Shown are the results from the simultation of the polarized (circles) and unpolarized (dots) UEG, where for the unpolarized case $2\cdot N_B$ spin-orbitals have been used.   
 }
 \label{fig:avgSign}
\end{figure}
The difference between CPIMC simulations of the polarized and unpolarized UEG enters basically in two ways. First, in addition to the particle number $N$, the total spin projection in the summation over the starting determinant $\{n^{(0)}\}$ in Eq.~(\ref{eq:Z_expansion}) has to be fixed, i.e., the number of spin-up $N_\uparrow$ and spin-down electrons $N_\downarrow$. Thus, if a whole occupied orbital is excited during the MC procedure (for details see Ref.~\cite{tim2}), it can only be excited to an orbital with the same spin projection. For example, orbital $6$ in Fig.~\ref{fig:sketch} could only be excited to orbital $8$ or some higher unoccupied orbital with spin up (not pictured). Moreover, when adding a kink or changing two kinks via some two-particle excitation, it is most effective to include spin conservation in the choice of the four involved orbitals, since all other proposed excitations would be rejected due to a vanishing weight.

For the second aspect, we have to explicitly consider the modulus weight of some kink $s_i=(pqrs)$, which is given by the modulus of Eq.~(\ref{eq:off_diagonal})
\begin{align}
&|Y_{ \{n^{(i)}\},\{n^{(i-1)}\} }(s_i)|=\nonumber\\
&\;\left| 
\frac{1}{(\mathbf{k}_{p} - \mathbf{k}_{r})^2}\delta_{\sigma_p,\sigma_r}\delta_{\sigma_q,\sigma_s}
-\frac{1}{(\mathbf{k}_{p} - \mathbf{k}_{s})^2}\delta_{\sigma_p,\sigma_s}\delta_{\sigma_q,\sigma_r}
\right|\nonumber\\
&\;\cdot\frac{4\pi e^2}{L^3}\delta_{\mathbf{k}_p+\mathbf{k}_q, \mathbf{k}_r + \mathbf{k}_s}\;,
\label{eq:spin_blocking}
\end{align}
where we have used the definition of the anti-symmetrized two-electron integrals from Sec.~\ref{sec:second_quant}. If all of the involved spin-orbitals have the same spin projection, the Kronecker deltas due to the spin obviously equal one, and the two-electron integrals are efficiently blocked, i.e., in most (momentum conserving) cases it is $|w^-_{pqrs}|<|w_{pqrs}|$ and $|w^-_{pqrs}|<|w_{pqsr}|$. However, if the involved orbitals have different spin projections, one of the two terms in Eq.~(\ref{eq:spin_blocking}) is always zero and $|w^-_{pqrs}|=|w_{pqsr}|$ or $|w^-_{pqrs}|=|w_{pqrs}|$. Hence, for otherwise fixed system parameters, the average weight of kinks in the unpolarized system is significantly larger. Since the diagonal matrix elements, cf.~Eq.~(\ref{eq:diagonal}), are independent of the spin, there ought to be more kinks in simulations of the unpolarized system, which in turn results in a smaller sign, because each kink enters the partition function with three possible sign changes. 

We address this issue in Fig.~\ref{fig:avgSign}, where we plot the average sign \textbf{a)} and the average number of kinks \textbf{b)} for the polarized (circles) and unpolarized (dots) UEG of $N=4,14$ and $66$ electrons at $\theta=1$. Coming from small values of $r_s$, the average number of kinks grows linearly with $r_s$. Depending on the particle number, at some critical value of $r_s$, it starts growing exponentially, until it eventually turns again into a linear dependency. The onset of the exponential growth is connected to a drop of the average sign due to the combinatorial growth of potential sign changes in the sampled paths with increasing number of kinks. This behavior becomes more extreme the larger the particle number, both for the polarized and unpolarized system, so that for $N=66$ electrons (blue lines), the average number of kinks suddenly increases from less than about two to a couple of hundred, which corresponds to a drop of the average sign from almost one to below $10^{-3}$. However, for the unpolarized system, the critical value of $r_s$ at which the average sign starts dropping drastically is approximately half of that of the polarized system containing the same number of electrons. In practice, this means that for $N=66$ polarized electrons at $\theta=1$ direct CPIMC calculations are feasible up to $r_s\sim 0.6$, whereas for $N=66$ unpolarizd electrons direct CPIMC is applicable only up to $r_s\sim 0.3$.

\begin{figure}[]
 \centering
 \includegraphics{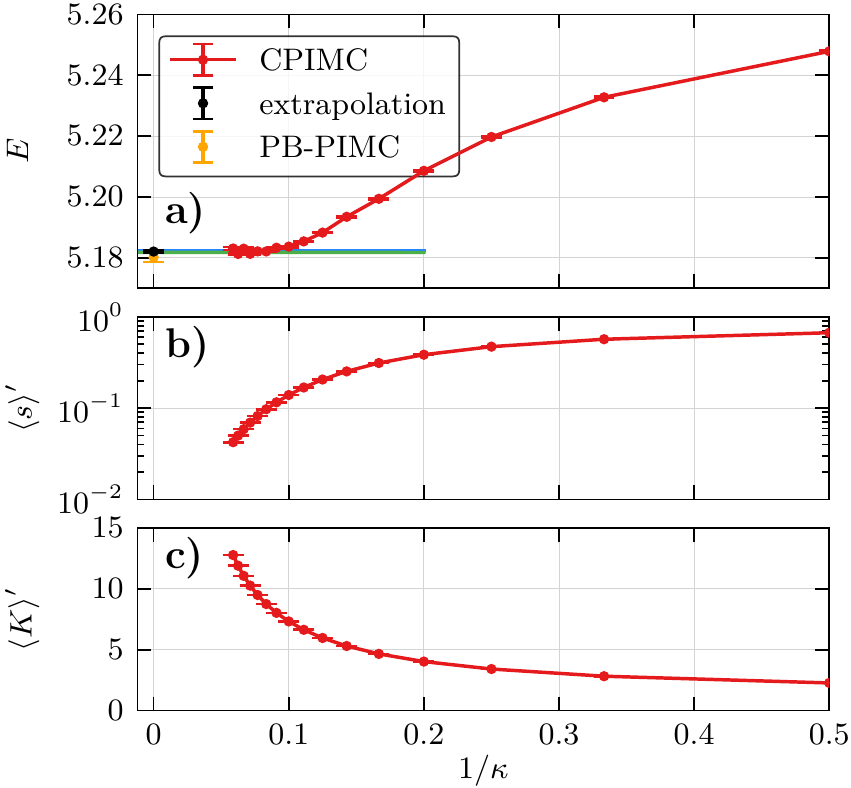}
 \caption{\label{convergence1}
 Convergence of \textbf{a)} the internal energy, \textbf{b)} the average sign and \textbf{c)} the average number of kinks with respect to the kink potential parameter $\kappa$ of $N=66$ unpolarized electrons at $r_s=2$ and $\theta=4$ in  $N_B=88946$ spin orbitals. The potenital parameter $\delta$ has been fixed to one. The blue (green) line show a horizontal (linear) fit to the last converged points. The asymptotic value (black point) in the limit $1/\kappa \to 0$ is enclosed between the blue and green lines and, within error bars, coincides with the PB-PIMC result (orange points).
 }
\end{figure}
\subsubsection{Auxiliary kink potential\label{kinkpot}}
In Ref.~\cite{groth}, it has been shown that the use of an auxiliary kink potential of the form
\begin{align}
V_{\delta,\kappa}(K)=\frac{1}{e^{-\delta(\kappa-K+0.5)}+1}\;
\label{eq:fermiPot}
\end{align}
significantly extends the applicability range of our CPIMC method towards larger values of $r_s$. This is achieved by adding the potential to the second line of the partition function Eq.~(\ref{eq:Z_expansion}), i.e., multiplying the weight of each path with the potential. Obviously, since $V_{\delta,\kappa}(K)\to 1$ in the limit $\kappa\to\infty$, performing CPIMC simulations for increasing values of $\kappa$ at fixed $\delta$ always converges to the exact result. Yet, to ensure a monotonic convergence of the energy, it turned out that the value of $\delta$ has to be sufficiently small. Both for the polarized and unpolarized system, choosing $\delta=1$ is sufficient. In fact, the potential is nothing but a smooth penalty for paths with a larger number of kinks than $\kappa$. 

In Fig.~\ref{convergence1}, we show the convergence of \textbf{a)} the internal energy (per particle), \textbf{b)} the average sign and \textbf{c)} the average number of kinks with respect to the kink potential parameter $\kappa$ of $N=66$ unpolarized electrons at $r_s=2$ and $\theta=4$. We have performed independent CPIMC simulations for different $\kappa$, using integer values from $2$ to $17$. While the energy almost remains constant for $\kappa\ge10$ with a corresponding average sign larger than $0.1$, the average sign and number of kinks themselves clearly are not converged. Further, the direct CPIMC algorithm (without the kink potential) would give a couple of thousand kinks with a practically vanishing sign. However, for the convergence of observables like the energy, apparently, a significantly smaller number of kinks is sufficient. This can be explained by a near cancellation of all additional contributions of the sampled paths with increasing number of kinks. For a detailed analysis, see Ref.~\cite{groth}.

We generally observe an s-shaped convergence of observables with $1/\kappa$, where the onset of the cancellation and near convergence are clearly indicated by the change in curvature. This allows for a robust extrapolation scheme to the asymptotic limit $1/\kappa\to\infty$, which is explained in detail in Ref.~\cite{groth}. An upper (lower) bound of the asymptotic value is obtained by a horizontal (linear) fit to the last points after the onset of convergence. The extrapolated result is then computed as the mean value of the lower and upper bounds with the uncertainty estimated as their difference. In Fig.~\ref{convergence1}, both, the horizontal (blue line) and linear fit (green line) almost coincide due to the complete convergence (within statistical errors) of the last points. The asymptotic CPIMC result (black dot) perfectly agrees (within error bars) with the PB-PIMC result (orange dot). This confirms the validity of using the kink potential also for the unpolarized UEG. 

\begin{figure}[]
 \centering
 \includegraphics{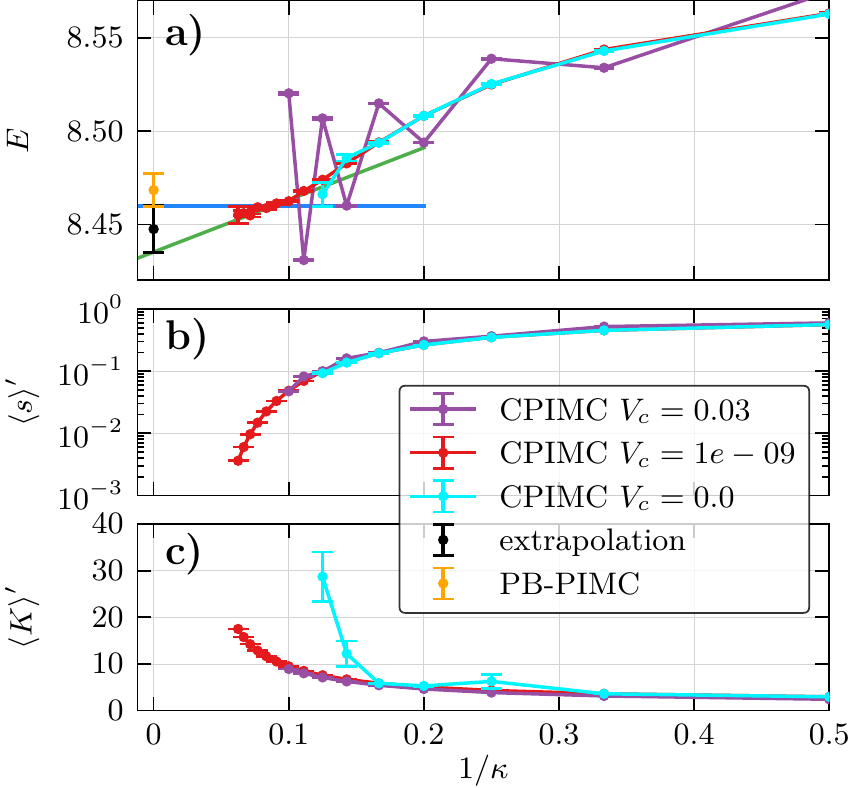}
 \caption{\label{convergence2}
 Convergence of \textbf{a)} the internal energy, \textbf{b)} the average sign and \textbf{c)} the average number of kinks with respect to the kink potential parameter $\kappa$ of $N=66$ unpolarized electrons at $r_s=0.8$ and $\theta=1$ in  $N_B=11150$ spin orbitals. The potential parameter $\delta$ has been fixed to one. The three curves correspond to CPIMC calculations where the kink potential has been cut off at different values $V_c$, i.e., $V_{1,\kappa}(K)$ (cf.~Eq.~(\ref{eq:fermiPot})) is set to zero if it takes values smaller than $V_c$.
 The blue (green) line shows a horizontal (linear) fit to the last converged red points. The asymptotic value (black point) in the limit $1/\kappa \to 0$ is enclosed between the blue and green lines and, within error bars, coincides with the PB-PIMC result (orange points).
 }
\end{figure}
\subsubsection{Further enhancement of the kink potential}
It turns out that, in case of the unpolarized UEG, even with the use of a kink potential with $\delta=1$, the simulation may approach paths with an extremely large number of kinks. This is demonstrated by the turquoise data points in Fig.~\ref{convergence2}~\textbf{c}), where the average number of kinks is shown for $N=66$ unpolarized electrons at $\theta=1$ and $r_s=0.8$. For example, at $\kappa=8$, there are on average about $30$ kinks. However, increasing the penalty for paths with a number of kinks larger than $\kappa$, by increasing $\delta$, is not a solution, since this would cause a non-monotonic convergence, oscillating with even and odd numbers of $\kappa$, as has been demonstrated in Ref.~\cite{groth}. Therefore, we choose a different strategy which is justified by the fact that paths with a very large number of kinks do not contribute to physical observables, cf. Sec.~\ref{kinkpot} and Ref.~\cite{groth}: we cut off the potential once it has dropped below some critical value $V_c$, thereby completely prohibiting paths where $V_{1,\kappa}(K)<V_c$. If the cut-off value is too large, we again recover an oscillating convergence behavior of the energy with even and odd numbers of $\kappa$ rendering an extrapolation difficult. This is shown by the purple data points in Fig.~\ref{convergence2}~\textbf{a)}, where the simulations have been performed with $V_c=0.03$ so that paths with a number of kinks larger than $\kappa+3$ are prohibited. On the other hand, if we set $V_c=10^{-9}$, so that paths with up to $\kappa+20$ kinks are allowed, the oscillations vanish (within statistical errors) and we can again apply our extrapolation scheme. Indeed, even with the additional cut-off the extrapolated value (black dot) coincides with that of the PB-PIMC simulation (orange dot) within error bars. In all simulations presented below we have carefully verified that the cut-off value is sufficiently small to guarantee converged results.

To summarize, as for the polarized UEG \cite{groth}, the accessible range of density parameters $r_s$ of our CPIMC method can be extended by more than a factor two by the use of a suitable kink potential, in simulations of the unpolarized UEG as well. For example, at $\theta=1$ direct CPIMC simulations are feasible up to $r_s\sim 0.3$, see Fig.~\ref{fig:avgSign}, whereas the kink potential allows us to obtain accurate energies up to $r_s=0.8$, as demonstrated in Fig.~\ref{convergence2}. In addition to the extrapolation scheme that has been introduced before for the spin-polarized case~\cite{groth}, we have cut off the potential at a sufficiently small value to prevent the simulation paths from approaching extremely large numbers of kinks. We expect this enhancement of CPIMC to be useful for arbitrary systems. In particular, it will allow us to further extend our previous results for the polarized UEG to larger $r_s$-values.

\section{Combined CPIMC and PB-PIMC Results\label{combined}}
\subsection{Exchange correlation energy\label{combined_exc}}

\begin{figure}[]
 \centering
 \includegraphics{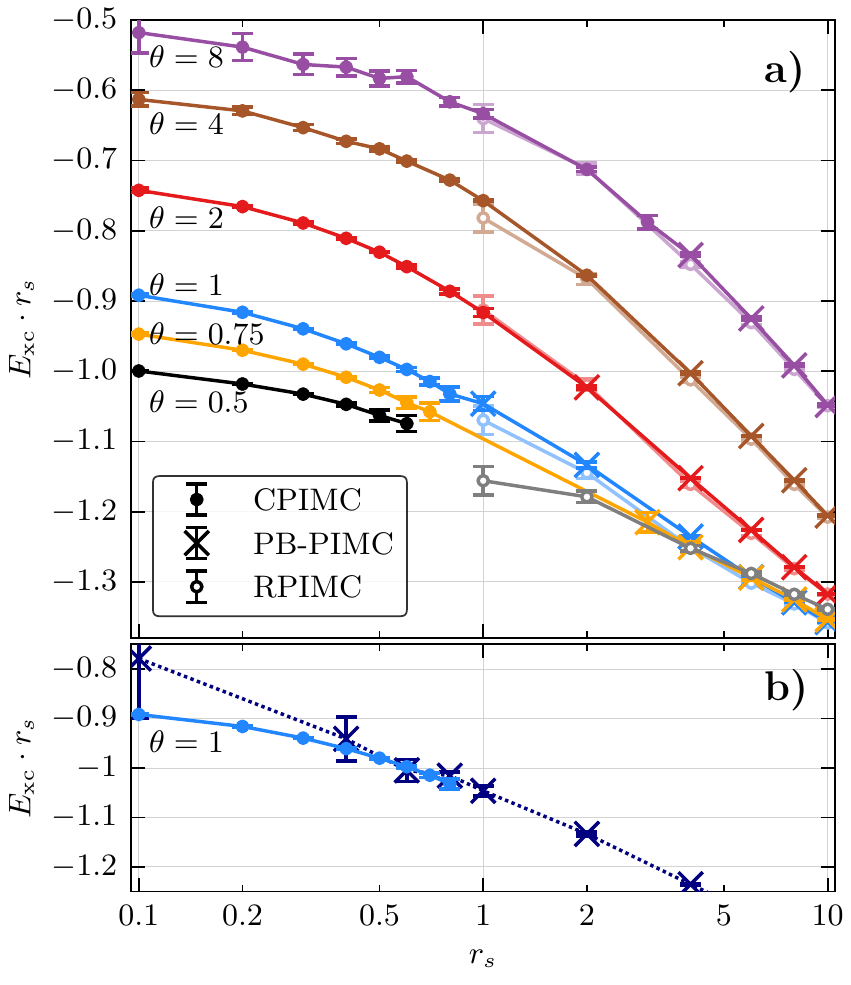}
 \caption{\label{Exc}
 Exchange-correlation energy $E_{xc}$ times $r_s$ of the unpolarized $N=66$ particle UEG over the density parameter $r_s$ for different temperatures. In graphic \textbf{a)}, only the best results from CPIMC (dots) or PB-PIMC (crosses) calculations are shown, cf.~table \ref{tab:1} in the appendix. In addition, RPIMC results by Brown \textit{et al.}\cite{brown,brown2} are plotted for comparison (lines with light colors and open circles). Graphic \textbf{b)} also shows PB-PIMC data for $r_s<1$ at $\theta=1$.}
\end{figure}
The exchange-correlation energy per particle, $E_{xc}$, of the uniform electrons gas is of central importance for the construction of density functionals and, therefore, has been the subject of numerous previous studies, e.g., Refs.~\cite{brown,brown3,karasiev,vfil4,prl,groth}. It is defined as the difference between the total energy of the correlated system and the ideal energy $U_0$,
\begin{eqnarray}
E_{xc} = E - U_0 \; . \label{def_exc}
\end{eqnarray}
In Fig.~\ref{Exc} a), we show results for this quantity for six different temperatures in dependence on the density parameter $r_s$. All data are also available in Tab.~\ref{tab:1} in the appendix. In order to fully exploit the complementary nature of our two approaches, we always present the most accurate data from either CPIMC (dots) or PB-PIMC (crosses). This allows us to cover the entire density range for $\theta \ge 1$, since here, the two methods allow for an overlap with respect to $r_s$. For completeness, we mention that the apparently larger statistical uncertainty for $\theta=8$ in comparison to lower temperature is not a peculiar manifestation of the FSP, but, instead, an artifact due to the definition~(\ref{def_exc}). At high temperature, the system becomes increasingly ideal and, therefore, the total energy $E$ approaches $U_0$. To obtain $E_{xc}$ at $\theta=8$, a large part of $E$ is subtracted, which, obviously, means that the comparatively small remainder is afflicted with a larger statistical uncertainty.  

To illustrate the overlap between PB-PIMC and CPIMC, we show all available data points for $\theta=1$ for both methods in panel b). This is the lowest temperature for which this is possible and, therefore, the most difficult example, because the systematic propagator error from PB-PIMC at small $r_s$ is most significant here. Evidently, both data sets are in excellent agreement with each other and the deviations are well within the error bars. Although we do expect that the deterioration of the convergence of the PB-PIMC factorization scheme for small $r_s$, cf.~Fig.~\ref{error}, should become less severe for larger systems, any systematic trend is masked by the sign problem anyway and cannot clearly be resolved for the given statistical uncertainty. 

Let us now consider temperatuers below $\theta=1$. For $\theta=0.75$, CPIMC is applicable only for $r_s\le 0.7$, while PB-PIMC delivers accurate results for $r_s\ge3$. Thus, the intermediate regime remains, without further improvements, out of reach and, for $\theta=0.5$, PB-PIMC is not applicable for $N=66$ unpolarized electrons in this density regime at all.

The comparison of our new combined results to the RPIMC data by Brown \textit{et al}.~\cite{brown}, which are available for $r_s\ge 1$, reveals excellent agreement for the three highest temperatures, $\theta=2,4,8$. For $\theta=1$, all results are still within single error bars, but the RPIMC data appear to be systematically too low. This observation is confirmed for $\theta=0.5$, where the fixed node approximation seems to induce an even more significant drop of $E_{xc}$. For completeness, we mention that although a similar trend has been found for the spin-polarized UEG as well \cite{prl,dornheim2,groth}, the overall agreement between RPIMC and our independent results is a little better for the unpolarized case.

\begin{figure}[]
 \centering
 \includegraphics{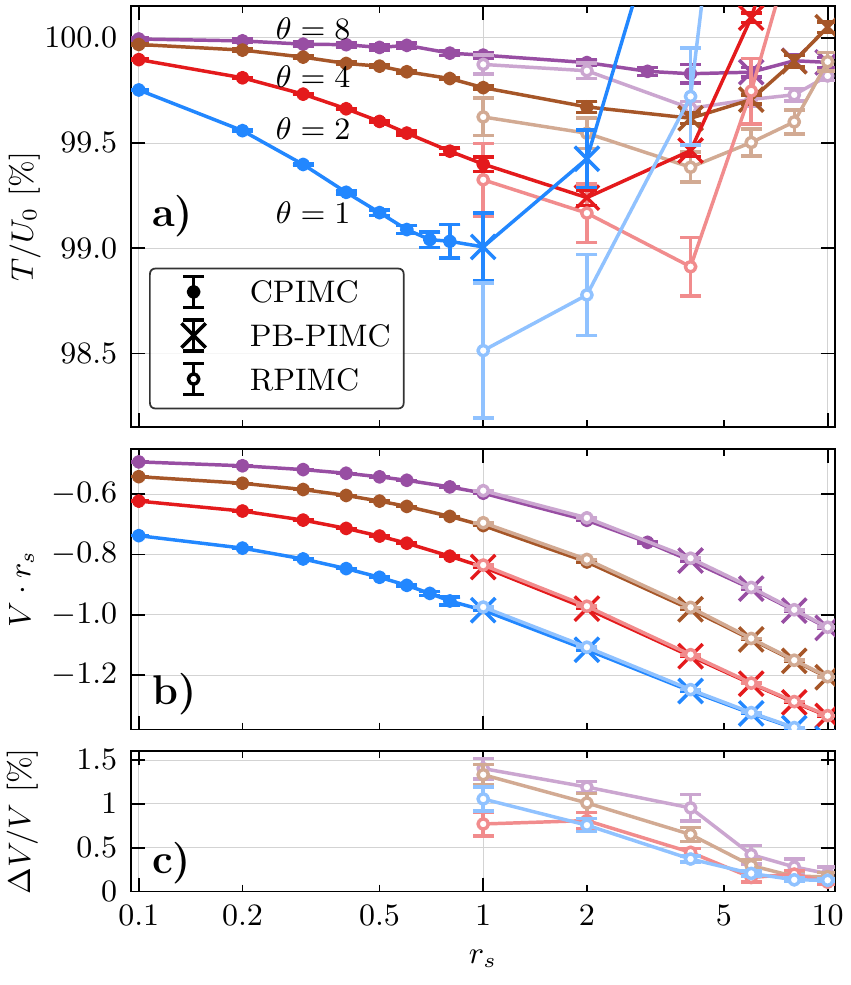}
 \caption{\label{W_and_K}
 Kinetic (a) and potential (b) energy of the unpolarized $N=66$ particle UEG over the density parameter $r_s$ for different temperatures.
  Panel c) shows the relative difference between our results and RPIMC data by Brown \textit{et al.}~\cite{brown,brown2}.}
\end{figure}
Finally, we consider the kinetic and potential contribution, $K$ and $V$, to the total energy separately.
In Fig.~\ref{W_and_K} a), the kinetic energy in units of the ideal energy $U_0$ is plotted versus $r_s$ and we again observe excellent agreement between PB-PIMC and CPIMC for all four shown temperatures.
The RPIMC data, on the other hand, exhibit clear deviations and are systematically too low even for $r_s=10$. In panel b), we show the same information for the potential energy, but the large $V$-range prevents us from resolving any differences between the different data sets.
For this reason, in panel c), we explicitly show the relative differences between our new results and those from RPIMC. Evidently, the latter are systematically too high and the relative deviations increase with density exceeding $\Delta V/V=1\%$. Curiously, $\Delta V/V$ attains its largest value for the highest temperature, $\theta=8$, which contradicts the usual assumption that the nodal error decreases with increasing $\theta$. Yet, in case of the exchange correlation energy, cf. Fig.~\ref{Exc}, this trend seems to hold. 

We summarize that, while RPIMC exhibits significant deviations for both $K$ and $V$ separately, these almost exactly cancel and, therefore, the total energy (and $E_{xc}$) is in rather good agreement with our results. This trend is in agreement with previous observations for the spin-polarized case \cite{dornheim2}.

\subsection{Pair distribution function\label{combined_pair}}

  \begin{figure}[]
 \centering
 \includegraphics[width=0.495\textwidth]{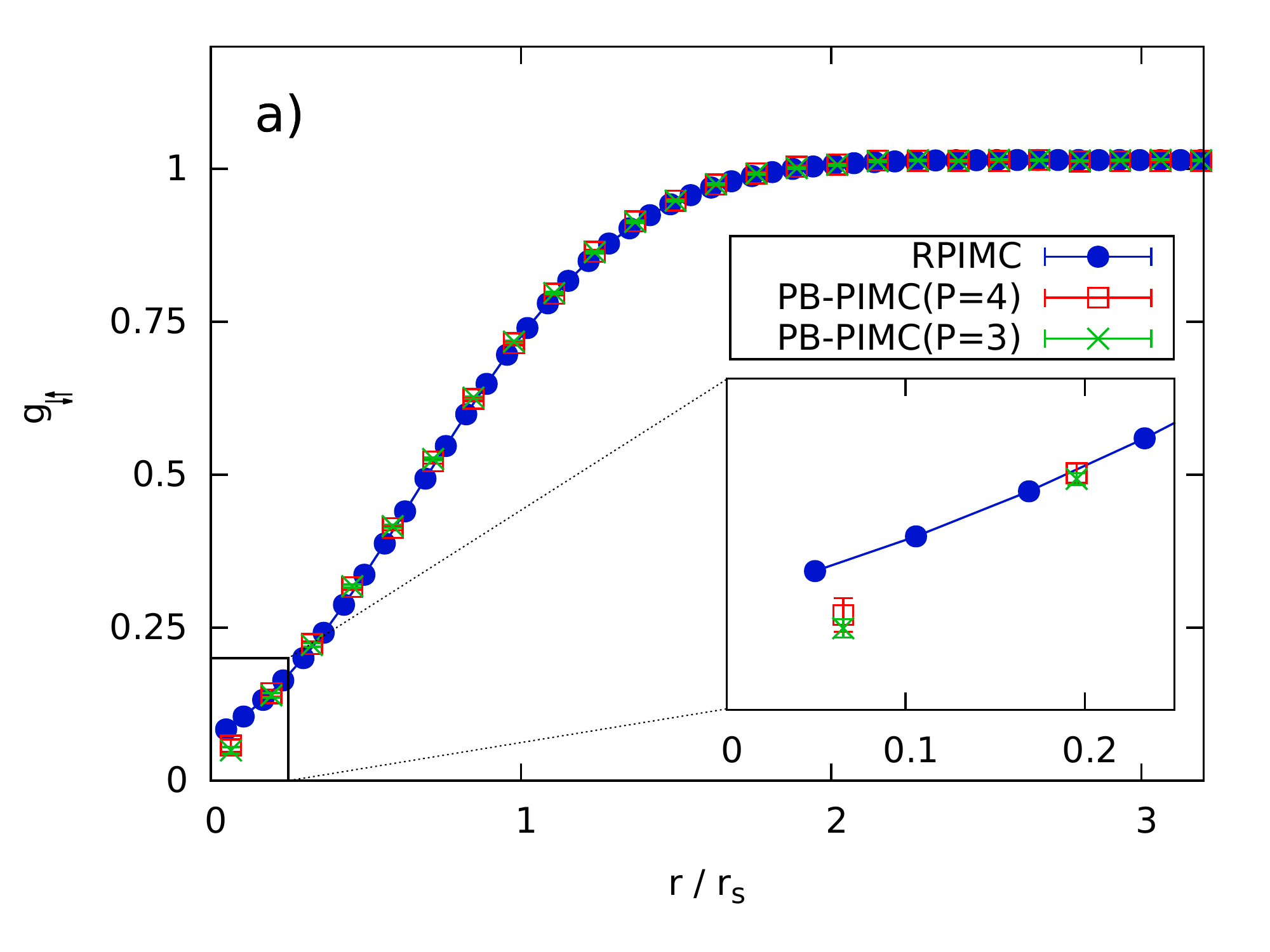}
  \includegraphics[width=0.495\textwidth]{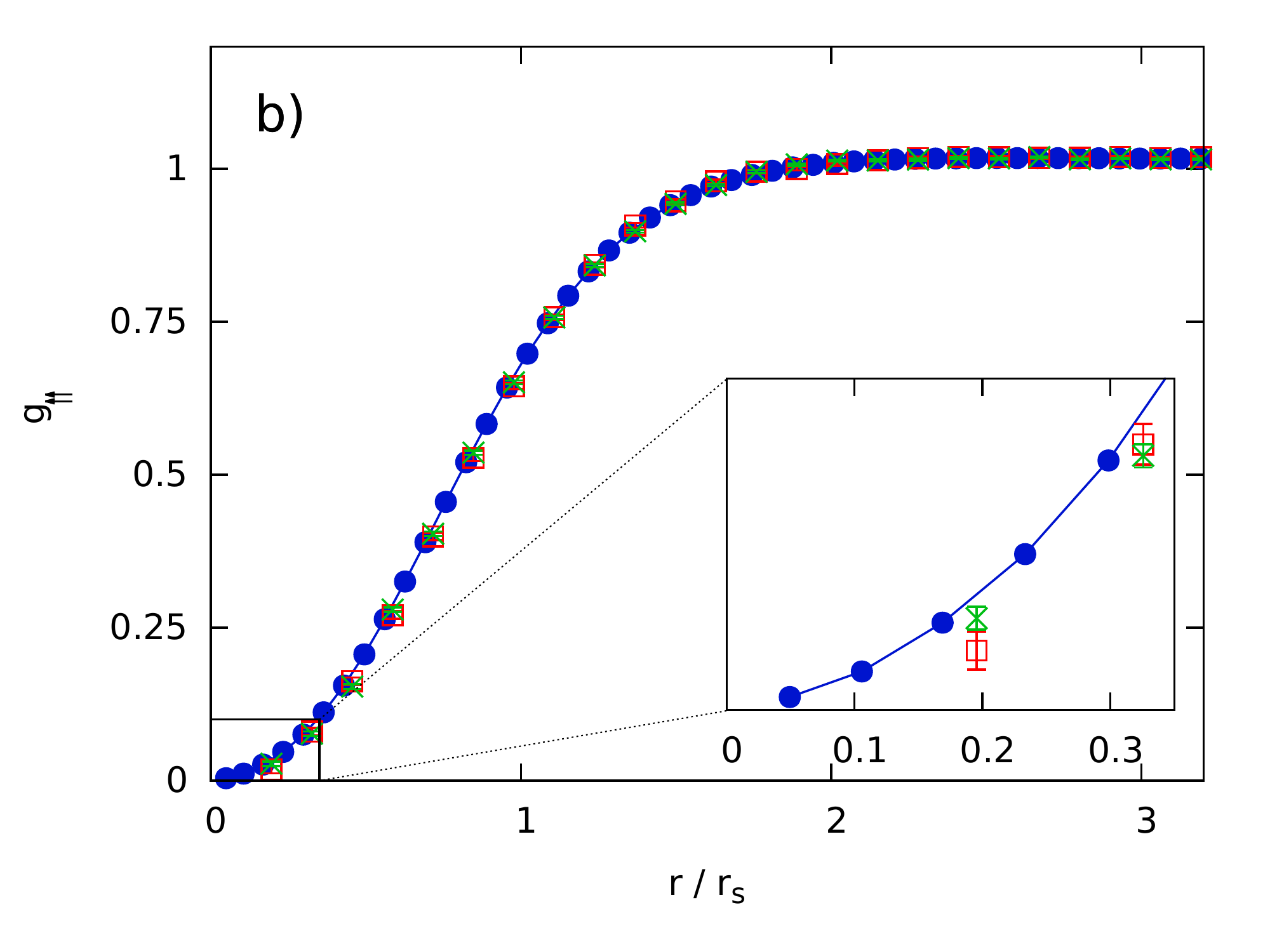}
 \caption{\label{brown_cf}Pair distribution function of $N=66$ unpolarized electrons at $r_s=4$ and $\theta=1$  -- The PB-PIMC results have been obtained for $t_0=0.04$ and $a_1=0$, and the RPIMC data are taken from Ref.~\cite{brown}.
 }
\end{figure}
Up to this point, we have compared RPIMC data for various energies ($E_{xc}$, $V$, $K$) to our independent results. However, since only the total energy was in agreement while $V$ and $K$ both deviated, it remains an open question how other thermodynamic quantities are affected by the fixed node approximation.
To address this issue, in Fig.~\ref{brown_cf} we show results for the pair distribution function (PDF) of the $N=66$ unpolarized electrons at $r_s=4$ and $\theta=1$. This appears to be the most convenient parameter combination for a comparison since, on the one hand, there are significant differences for both $K$ and $V$ while, on the other hand, simulations with PB-PIMC are possible up to $P=4$, which allows for accurate results of both $g_{\uparrow\uparrow}$ and $g_{\uparrow\downarrow}$.
In panel a), the inter-species PDF $g_{\uparrow\downarrow}$ is plotted versus $r$ and shown are PB-PIMC results for $P=3$ (green crosses) and $P=4$ (red squares) as well as RPIMC data (blue circles) from Ref.~\cite{brown}. All three curves agree rather well and exhibit a distinct exchange correlation hole for $r\le 1.5 r_s$ and a featureless approach to unity at larger distances. The inset shows the short range part of the PDF, which is the only segment where deviations are visible. The PB-PIMC results for $P=3$ and $P=4$ are within each others error bars and, for the smallest resolved $r$, slightly below the RPIMC data, altough this trend hardly exceeds twice the error bars as well.
The results for the intra-species PDF $g_{\uparrow\uparrow}$ show a similar picture, although short range configurations of two particles are even more suppressed due to the Pauli blocking. Again, there appears a slight difference between PB-PIMC and RPIMC, which, however, cannot clearly be resolved within the given statistical uncertainty. 
Therefore, we conclude that our independent simulation data are in good agreement with the fixed node approximation for both pair distribution functions despite the observed deviations in $K$ and $V$ for these particular system parameters.

\section{Discussion}
In summary, we have successfully extended the combination of PB-PIMC and CPIMC, presented in paper I, to the unpolarized UEG and, thereby, presented unbiased \emph{ab initio} results at finite temperature. For PB-PIMC, we have observed an increased propagator error at high density, i.e., at $r_s<1$, compared to the polarized UEG. This issue arises from the absence of the Pauli blocking between electrons of different spin-polarization, making the combination with the complementary CPIMC approach indispensable.

On the other hand, CPIMC suffers from a significantly more severe FSP due to the increased configuration weight of inter-species kinks. To overcome this problem, we have developed an additional enhancement of our extrapolation scheme. The introduction of a (very small) cut-off parameter $V_c$ in the auxiliary kink potential prevents the number of kinks from diverging and, thereby, significantly extends the parameter range where accurate simulations are feasible.
We have demonstrated that CPIMC and PB-PIMC reveal excellent agreement, where both are available, and, in their combination, allow for accurate results over the entire density range, for $\theta\ge1$ and $N=66$ electrons. 

Overall, the existing RPIMC data for the exchange correlation energy are in better agreement with our \emph{ab initio} results than for the spin-polarized UEG, but there seems to be a similar unphysical systematic drop around $r_s=1$ at low temperatures. Interestingly, the separate kinetic and potential contributions to the energy substantially deviate from our results by more than one percent.
Furthermore, for the first time, we have presented a comparison of the pair distribution functions $g_{\uparrow\uparrow}(r)$ and $g_{\uparrow\downarrow}(r)$, which are in good agreement with RPIMC .

It remains an important issue of future work to perform an extrapolation to the macroscopic limit, i.e., the development of finite-size corrections, e.g., \cite{fraser,lin,drummond}. To this end simulations with substantially larger particle numbers are required which should be possible with the presented enhancements.
Furthermore, we expect that the presented combination of the complementary CPIMC and PB-PIMC approaches can be successfully applied to numerous other Fermi systems, such as two-component plasmas \cite{bonitz,morales,proton} and atoms embedded in jellium \cite{at1,at2,at3}.

\section*{Acknowledgements}
This work is supported by the Deutsche Forschungsgemeinschaft via project BO 1366-10 and via SFB TR-24 project A9 as well as grant shp00015 for CPU time at the Norddeutscher Verbund f\"ur Hoch- und H\"ochstleistungsrechnen (HLRN).

\section*{Appendix}
As a supplement to Figs.~\ref{Exc} and~\ref{W_and_K}, we have listed all combined simulation data from PB-PIMC and CPIMC in Tab.~\ref{tab:1}.

\begin{longtable*}[c]{@{}rr
@{\hspace{15pt}}  
S[table-parse-only,  table-text-alignment = center, table-number-alignment = left]
@{\hspace{15pt}}                                                              
S[table-parse-only,  table-text-alignment = center, table-number-alignment = left]
@{\hspace{15pt}}                                                              
S[table-parse-only,  table-text-alignment = center, table-number-alignment = left]
@{\hspace{15pt}}                                                              
S[table-parse-only,  table-text-alignment = center, table-number-alignment = left]
@{\hspace{15pt}}                                                              
S[table-parse-only,  table-text-alignment = center, table-number-alignment = left]
@{}}
\caption{Energies per particle for $N=66$ unpolarized electrons: ideal energy, $U_0$, kinetic energy, $T$, potential energy, $V$ and exchange-correlation energy $E_\text{xc}$. While the unmarked results correspond to standard CPIMC simulations (without the auxiliary kink potential), the \enquote{a} marks CPIMC results that have been obtained by the extrapolation as explained in Sec.~\ref{kinkpot} and Ref.~\cite{groth}. For the latter values, the error includes systematic effects. All other errors correspond to a $1\sigma$ standard deviation. A \enquote{b} marks results from PB-PIMC calculations. For CPIMC results, the utilized number of basis functions $N_B$ is given in the last column and has been fixed for the same temperature. The ideal energies have been computed using the same number of basis functions as for the interacting system. Energies in units of Ryd.}
\label{tab:1}\\\\
\toprule
 {$\theta$} &     {$r_s$} &              {$U_0$} &                          {$E_\text{kin}$} &                          {$E_\text{pot}$} &                        {$E_\text{xc}$} &  {$N_B$}\\
\midrule
\endfirsthead
\caption[]{\textit{(continued).} 
}\\\\
\toprule
 {$\theta$} &     {$r_s$} &              {$U_0$} &                          {$T$} &                          {$V$} &                        {$E_\text{xc}$} &  {$N_B$} \\
\midrule\endhead
 0.50 &   0.1 &    374.8592(12) &    373.463(6){${}^{}$} &   -8.60129(19){${}^{}$} &      -9.997(6){${}^{}$} &   11150 \\
  0.50 &   0.2 &    93.71481(30) &   93.1294(25){${}^{}$} &      -4.506(4){${}^{}$} &    -5.0911(25){${}^{}$} &   11150 \\
  0.50 &   0.3 &    41.65102(13) &   41.3226(28){${}^a$} &    -3.1130(10){${}^a$} &     -3.4421(9){${}^a$} &   11150 \\
  0.50 &   0.4 &     23.42870(8) &   23.2220(29){${}^a$} &      -2.409(4){${}^a$} &      -2.618(6){${}^a$} &   11150 \\
  0.50 &   0.5 &     14.99437(5) &    14.871(18){${}^a$} &     -1.992(20){${}^a$} &     -2.126(16){${}^a$} &   11150 \\
  0.50 &   0.6 &   10.412756(34) &    10.327(15){${}^a$} &     -1.702(33){${}^a$} &     -1.791(19){${}^a$} &   11150 \\
  \midrule
  \pagebreak[3]
  0.75 &   0.1 &      495.690(4) &   494.119(16){${}^{}$} &   -7.90080(19){${}^{}$} &     -9.472(17){${}^{}$} &   11150 \\
  0.75 &   0.2 &    123.9225(10) &  123.2322(29){${}^{}$} &   -4.16057(12){${}^{}$} &    -4.8508(31){${}^{}$} &   11150 \\
  0.75 &   0.3 &      55.0767(5) &     54.672(4){${}^a$} &   -2.89413(31){${}^a$} &    -3.2999(14){${}^a$} &   11150 \\
  0.75 &   0.4 &    30.98062(26) &     30.712(4){${}^a$} &    -2.2506(18){${}^a$} &    -2.5215(30){${}^a$} &   11150 \\
  0.75 &   0.5 &    19.82760(17) &     19.637(4){${}^a$} &      -1.858(5){${}^a$} &      -2.054(8){${}^a$} &   11150 \\
  0.75 &   0.6 &    13.76916(12) &    13.632(10){${}^a$} &     -1.601(17){${}^a$} &     -1.741(14){${}^a$} &   11150 \\
  0.75 &   0.7 &     10.11612(9) &    10.018(18){${}^a$} &     -1.400(23){${}^a$} &     -1.511(18){${}^a$} &   11150 \\
  0.75 &   3.0 &     0.550767(5) &      0.556(5){${}^b$} &     -0.4098(8){${}^b$} &      -0.405(5){${}^b$} &         \\
  0.75 &   4.0 &   0.3098060(26) &    0.3173(18){${}^b$} &     -0.3201(4){${}^b$} &    -0.3127(18){${}^b$} &         \\
  0.75 &   6.0 &   0.1376920(12) &     0.1469(6){${}^b$} &   -0.22488(13){${}^b$} &     -0.2157(5){${}^b$} &         \\
  0.75 &   8.0 &    0.0774520(7) &   0.08610(19){${}^b$} &    -0.17428(6){${}^b$} &   -0.16563(19){${}^b$} &         \\
  0.75 &  10.0 &    0.0495690(4) &    0.05687(9){${}^b$} &  -0.142666(28){${}^b$} &    -0.13536(9){${}^b$} &         \\
  \midrule
  \pagebreak[3]
  1.00 &   0.1 &      623.230(6) &   621.686(15){${}^{}$} &    -7.37511(9){${}^{}$} &     -8.918(17){${}^{}$} &   11150 \\
  1.00 &   0.2 &    155.8074(15) &  155.1203(34){${}^{}$} &   -3.89359(12){${}^{}$} &      -4.581(4){${}^{}$} &   11150 \\
  1.00 &   0.3 &      69.2477(7) &   68.8312(18){${}^{}$} &   -2.71561(11){${}^{}$} &    -3.1322(19){${}^{}$} &   11150 \\
  1.00 &   0.4 &      38.9518(4) &   38.6661(33){${}^a$} &     -2.1165(8){${}^a$} &    -2.4025(25){${}^a$} &   11150 \\
  1.00 &   0.5 &    24.92918(24) &   24.7222(32){${}^a$} &    -1.7508(17){${}^a$} &      -1.961(4){${}^a$} &   11150 \\
  1.00 &   0.6 &    17.31193(17) &   17.1543(34){${}^a$} &      -1.503(4){${}^a$} &      -1.663(4){${}^a$} &   11150 \\
  1.00 &   0.7 &    12.71897(12) &     12.597(5){${}^a$} &     -1.327(10){${}^a$} &      -1.450(7){${}^a$} &   11150 \\
  1.00 &   0.8 &      9.73796(9) &      9.644(8){${}^a$} &     -1.192(16){${}^a$} &     -1.290(13){${}^a$} &   11150 \\
  1.00 &   1.0 &      6.23230(6) &     6.170(10){${}^b$} &    -0.9844(10){${}^b$} &     -1.046(10){${}^b$} &         \\
  1.00 &   2.0 &    1.558074(15) &    1.5491(21){${}^b$} &   -0.55777(28){${}^b$} &    -0.5667(21){${}^b$} &         \\
  1.00 &   4.0 &     0.389518(4) &   0.39370(21){${}^b$} &    -0.31304(5){${}^b$} &   -0.30886(21){${}^b$} &         \\
  1.00 &   6.0 &   0.1731190(17) &   0.17863(15){${}^b$} &    -0.22107(4){${}^b$} &   -0.21556(15){${}^b$} &         \\
  1.00 &   8.0 &    0.0973800(9) &    0.10313(6){${}^b$} &  -0.171900(18){${}^b$} &    -0.16615(6){${}^b$} &         \\
  1.00 &  10.0 &    0.0623230(6) &  0.067639(31){${}^b$} &  -0.141041(11){${}^b$} &  -0.135725(31){${}^b$} &         \\
  \midrule
  \pagebreak[3]
  2.00 &   0.1 &    1155.227(11) &  1154.031(32){${}^{}$} &   -6.22959(19){${}^{}$} &     -7.425(33){${}^{}$} &   18342 \\
  2.00 &   0.2 &    288.8066(28) &    288.258(7){${}^{}$} &    -3.27971(9){${}^{}$} &      -3.828(7){${}^{}$} &   18342 \\
  2.00 &   0.3 &    128.3585(12) &  128.0151(35){${}^{}$} &    -2.28648(6){${}^{}$} &      -2.630(4){${}^{}$} &   18342 \\
  2.00 &   0.4 &      72.2017(7) &   71.9583(17){${}^{}$} &    -1.78368(6){${}^{}$} &    -2.0270(18){${}^{}$} &   18342 \\
  2.00 &   0.5 &      46.2091(4) &   46.0256(11){${}^{}$} &    -1.47771(6){${}^{}$} &    -1.6612(11){${}^{}$} &   18342 \\
  2.00 &   0.6 &    32.08963(31) &   31.9444(29){${}^a$} &   -1.27090(35){${}^a$} &      -1.419(4){${}^a$} &   18342 \\
  2.00 &   0.8 &    18.05042(17) &   17.9532(27){${}^a$} &    -1.0069(11){${}^a$} &      -1.108(4){${}^a$} &   18342 \\
  2.00 &   1.0 &    11.55227(11) &     11.483(4){${}^a$} &    -0.8440(32){${}^a$} &      -0.916(5){${}^a$} &   18342 \\
  2.00 &   2.0 &    2.888066(28) &    2.8661(11){${}^b$} &   -0.48960(21){${}^b$} &    -0.5115(11){${}^b$} &         \\
  2.00 &   4.0 &     0.722017(7) &   0.71815(19){${}^b$} &    -0.28421(6){${}^b$} &   -0.28807(20){${}^b$} &         \\
  2.00 &   6.0 &   0.3208960(31) &    0.32120(7){${}^b$} &  -0.204649(24){${}^b$} &    -0.20434(8){${}^b$} &         \\
  2.00 &   8.0 &   0.1805040(17) &    0.18183(4){${}^b$} &  -0.161212(15){${}^b$} &    -0.15989(4){${}^b$} &         \\
  2.00 &  10.0 &   0.1155230(11) &  0.117282(28){${}^b$} &  -0.133507(13){${}^b$} &  -0.131748(32){${}^b$} &         \\
  \midrule
  \pagebreak[3]
  4.00 &   0.1 &    2245.508(30) &    2244.80(9){${}^{}$} &   -5.42045(19){${}^{}$} &      -6.13(10){${}^{}$} &   88946 \\
  4.00 &   0.2 &      561.377(8) &   561.050(26){${}^{}$} &    -2.81969(9){${}^{}$} &     -3.147(27){${}^{}$} &   88946 \\
  4.00 &   0.3 &    249.5008(34) &   249.272(14){${}^{}$} &    -1.94887(8){${}^{}$} &     -2.177(15){${}^{}$} &   88946 \\
  4.00 &   0.4 &    140.3442(19) &    140.173(8){${}^{}$} &    -1.51066(7){${}^{}$} &      -1.682(8){${}^{}$} &   88946 \\
  4.00 &   0.5 &     89.8203(12) &     89.699(6){${}^{}$} &    -1.24591(7){${}^{}$} &      -1.367(6){${}^{}$} &   88946 \\
  4.00 &   0.6 &      62.3752(8) &     62.275(4){${}^{}$} &    -1.06761(6){${}^{}$} &      -1.168(4){${}^{}$} &   88946 \\
  4.00 &   0.8 &      35.0861(5) &   35.0182(19){${}^{}$} &    -0.84205(6){${}^{}$} &    -0.9099(19){${}^{}$} &   88946 \\
  4.00 &   1.0 &    22.45508(30) &   22.4019(15){${}^{}$} &    -0.70405(7){${}^{}$} &    -0.7572(16){${}^{}$} &   88946 \\
  4.00 &   2.0 &      5.61377(8) &    5.5953(15){${}^a$} &   -0.41230(33){${}^a$} &     -0.4317(4){${}^a$} &   88946 \\
  4.00 &   4.0 &    1.403442(19) &     1.3981(4){${}^b$} &   -0.24535(17){${}^b$} &     -0.2507(4){${}^b$} &         \\
  4.00 &   6.0 &     0.623752(8) &   0.62192(14){${}^b$} &    -0.18022(7){${}^b$} &   -0.18205(16){${}^b$} &         \\
  4.00 &   8.0 &     0.350861(5) &    0.35047(9){${}^b$} &    -0.14402(4){${}^b$} &   -0.14441(11){${}^b$} &         \\
  4.00 &  10.0 &   0.2245510(30) &    0.22466(5){${}^b$} &  -0.120675(31){${}^b$} &    -0.12056(7){${}^b$} &         \\
  \midrule
  \pagebreak[3]
  8.00 &   0.1 &     4445.13(11) &   4444.88(27){${}^{}$} &   -4.93048(19){${}^{}$} &      -5.18(29){${}^{}$} &  147050 \\
  8.00 &   0.2 &    1111.281(27) &    1111.12(9){${}^{}$} &   -2.52994(12){${}^{}$} &      -2.69(10){${}^{}$} &  147050 \\
  8.00 &   0.3 &     493.903(12) &     493.75(5){${}^{}$} &    -1.72864(9){${}^{}$} &       -1.88(5){${}^{}$} &  147050 \\
  8.00 &   0.4 &      277.820(7) &   277.730(30){${}^{}$} &    -1.32690(8){${}^{}$} &     -1.417(31){${}^{}$} &  147050 \\
  8.00 &   0.5 &      177.805(4) &   177.724(22){${}^{}$} &    -1.08505(7){${}^{}$} &     -1.166(22){${}^{}$} &  147050 \\
  8.00 &   0.6 &    123.4757(30) &   123.431(15){${}^{}$} &    -0.92338(6){${}^{}$} &     -0.968(15){${}^{}$} &  147050 \\
  8.00 &   0.8 &     69.4551(17) &     69.404(7){${}^{}$} &    -0.71997(5){${}^{}$} &      -0.771(8){${}^{}$} &  147050 \\
  8.00 &   1.0 &     44.4513(11) &     44.415(6){${}^{}$} &    -0.59679(5){${}^{}$} &      -0.633(6){${}^{}$} &  147050 \\
  8.00 &   2.0 &    11.11281(27) &   11.0997(16){${}^{}$} &    -0.34329(5){${}^{}$} &    -0.3564(16){${}^{}$} &  147050 \\
  8.00 &   3.0 &     4.93903(12) &     4.9312(9){${}^a$} &     -0.2532(5){${}^a$} &    -0.2626(33){${}^a$} &  147050 \\
  8.00 &   4.0 &      2.77820(7) &     2.7746(6){${}^b$} &   -0.20502(29){${}^b$} &     -0.2086(6){${}^b$} &         \\
  8.00 &   6.0 &    1.234757(30) &   1.23274(28){${}^b$} &   -0.15214(15){${}^b$} &     -0.1542(4){${}^b$} &         \\
  8.00 &   8.0 &    0.694551(17) &   0.69379(18){${}^b$} &   -0.12321(10){${}^b$} &   -0.12396(23){${}^b$} &         \\
  8.00 &  10.0 &    0.444513(11) &   0.44399(11){${}^b$} &    -0.10430(7){${}^b$} &   -0.10482(13){${}^b$} &         \\
\bottomrule
\end{longtable*}

\end{document}